\documentclass[aps,prb,reprint,superscriptaddress]{revtex4-2}

\usepackage{gensymb}
\usepackage{placeins}

\usepackage{amsmath}
\usepackage{amsfonts}
\usepackage{amssymb}
\usepackage{relsize}
\usepackage{graphicx}
\usepackage{makecell}
\usepackage{float}
\usepackage{color}
\newcommand\lfrac[2]{\frac{\displaystyle #1}{\displaystyle #2}}

\begin{document}

\title{Interacting Two-Level Systems as a Source of 1/f Charge Noise in Quantum Dot Qubits}
\author{D. L. Mickelsen}
\author{Herv{\'e} M. Carruzzo}
\author{Clare C. Yu}
\affiliation{Department of Physics and Astronomy, University of California, Irvine, California 92697, USA}

\date{\today}

\begin{abstract}
Charge noise in semiconducting quantum dots has been observed to have a 1/f spectrum. We propose a model in which a pair of quantum dots are coupled to a 2D bath of fluctuating two level systems (TLS) that have electric dipole moments and that interact with each other, i.e., with the other fluctuators. These interactions are primarily via the elastic strain field. We use a 2D nearest-neighbor Ising spin glass to represent these elastic interactions and to simulate the dynamics of the bath of electric dipole fluctuators in the presence of a ground plane representing metal gates above the oxide layer containing the fluctuators. The interactions between the TLS cause the energy splitting of individual fluctuators to change with time. We calculate the resulting fluctuations in the electric potential at the two quantum dots that lie below the oxide layer. We find that 1/f electric potential noise spectra at the quantum dots and cross correlation in the noise between the two quantum dots are in qualitative agreement with experiment. Our simulations find that the cross correlations decrease exponentially with increasing quantum dot separation.
\end{abstract}

\newcommand{\be}{\begin{equation}}
\newcommand{\ee}{\end{equation}}
\newcommand{\beq}{\begin{eqnarray}}
\newcommand{\eeq}{\end{eqnarray}}
\maketitle

\section{Introduction}
Electron spins in Si/SiGe quantum dots (QDs) show promise as quantum bits, but are plagued by charge noise that goes as $1/f^\alpha$ over more than thirteen decades in frequency ~\cite{Yoneda2018,Connors2022}, where $f$ is the frequency and $\alpha$ is the noise exponent. 
1/f noise is typically attributed to a bath of charged 
two-level fluctuators with a broad distribution of switching rates.
If each fluctuator is associated with a double well
potential, then thermal activation over a uniform distribution of barrier 
heights produces 1/f noise \cite{Dutta1979,Dutta1981}.  With such a bath of two level systems (TLS), one would 
expect the charge noise to increase linearly with increasing
temperature \cite{Petit2020}. However, the measured temperature dependence
deviates from linearity 
\cite{Connors2019,Connors2019e,Petit2018,Spence2023}. In one case, the
deviation from linearity increases with increasing thickness of the gate
oxide layer \cite{Connors2019,Connors2019e} and, in another study, 
the temperature dependence of the charge noise
was found to be quadratic \cite{Petit2018}.

The small size of the quantum dot and the lack of temperature dependence 
between 0.45 K and 1.2 K in the decoherence time $T_2^*$ 
\cite{Petit2020} have led some to argue that each quantum dot is only coupled to a few fluctuators~\cite{Petit2020,Ahn2021,Throckmorton2023,Mickelsen2023}. This view was supported by the lack of correlation in the charge noise of neighboring quantum dots 
\cite{Connors2019,Connors2019e}, though
subsequent measurements on spin qubit quantum dots have found correlations between the charge noise in neighboring quantum dots that are about 100 nm apart \cite{Yoneda2023} and next-nearest neighbor quantum dots that are about 200 nm apart \cite{RojasArias2023}.  With only a few two-level fluctuators, a Lorentzian power spectra is expected instead of the observed 1/f noise.

A number of theoretical papers have considered the effect on semiconducting qubits of fluctuators that produce charge noise without going into the detailed microscopics of the fluctuators \cite{Hu2006,Culcer2009,Ramon2010,Li2010,Culcer2013,Huang2018,Yang2017,Shim2018}, though one group used the measured charge noise spectra to deduce the range of fluctuator rates \cite{Gungordu2019}. Those models that did consider specific types of fluctuators, e.g., charge fluctuators \cite{RojasArias2023,Shehata2023}, electron hopping \cite{Beaudoin2015,Shehata2023,Kepa2023,KepaCorrelations2023}, and fluctuators coupled to phonons \cite{Beaudoin2015}, had independent, noninteracting fluctuators.

It has been suggested that a few two-level fluctuators can produce 1/f noise if they are coupled to a microscopic subsection of the larger thermal bath \cite{Ahn2021,Throckmorton2023}.  Temperature fluctuations in a subsection of the 2D electron gas (2DEG) are proposed to cause 1/f noise over several decades of frequency, but this model requires that the temperature fluctuations be extremely slow, i.e., the sub-bath would need to remain at a single temperature for unphysically long periods of time in order to explain the low frequency noise~\cite{Mickelsen2023}.

In this paper, we propose that the charge noise arises from a bath of TLS with fluctuating electric dipole moments that reside outside the 2DEG, e.g., in the oxide layer \cite{King2020}. These TLS interact with one another via elastic and electric dipole moments, though the elastic interactions dominate. (As we show below, elastic dipole-dipole interactions are about an order of magnitude larger than electric dipole-dipole interactions \cite{Yu1985}.) The TLS energy level splittings change with time due to interactions with their fluctuating TLS neighbors. This bath of fluctuating electric dipole moments, together with their image charges in the ground plane lying below the metal gates, produces 1/f noise in the electric potential seen by the quantum dots (QDs) consistent with the observed 1/f charge noise \cite{Connors2022}. We find that the correlation of the noise on the quantum dots decreases exponentially with increasing separation between the QDs. When the QDs are about 100 nm apart, the noise is mildly correlated in agreement with experiment \cite{Yoneda2023}. The question of whether the noise in quantum dots is correlated is important because error correcting codes assume that the errors in qubits is independent and uncorrelated \cite{Fowler2012}. 

In the next section we describe our model of two quantum dots in a quantum well located below an oxide layer containing a 2D bath of TLS that have both elastic and electric dipole moments. A Monte Carlo simulation of a 2D nearest-neighbor Ising spin glass is used to represent the TLS interacting via the elastic strain field. When an Ising spin flips, this represents a flip of the elastic dipole moment and the associated electric dipole moment follows along by also flipping. The metallic gates that cover the surface of the device are treated as a ground plane with image charges corresponding to the images of the TLS electric dipole moments.  The fluctuating dipole moments and their images produce a fluctuating electric potential at the quantum dots that results in 1/f charge noise.  We go on to calculate the correlations in the noise at the two quantum dots. These results are presented in section III and discussed in section IV.

\section{Model}
In this section we describe our model that reflects 
the experimental setup of a pair of quantum dots that are a distance $d$
apart in a quantum well coupled to a bath of TLS with both elastic 
and electric dipole
moments residing in the oxide layer about 30 nm above the quantum well. A 
set of metal gates lies on top of the oxide layer and are represented by a
metal ground plane. The TLS electric dipole moments have image charges in the
ground plane; both the electric dipole moments and the image charges contribute
to the fluctuating electric potential at the quantum dots. The quantum dots
are passive, i.e., they have no internal dynamics; we merely use them to mark
the locations where we want to record the charge noise. To determine the
dependence of the correlation of the noise on the separation $d$ between
the QD, we set $d$ = 30, 50, 70, and 100 nm. 

Two level systems have long been used to describe the thermal and acoustic 
properties of amorphous materials at low temperatures 
\cite{Hunklinger1986,Phillips1987}.
The standard TLS model postulates the existence of independent entities 
that tunnel between the two minima of a double well potential with a 
flat distribution of tunnel barrier heights and energy asymmetries 
\cite{Anderson1972,Phillips1972}. Using a right-well left-well basis, 
the Hamiltonian representing a given TLS is 
\begin{equation}
H_{\rm TLS}=\frac{1}{2}\left(\Delta\sigma_z+\Delta_o\sigma_x\right),
\end{equation}
where $\sigma_x$ and $\sigma_z$ are Pauli matrices,
$\Delta$ is the TLS asymmetry energy, i.e., the energy difference between
the right and left wells, and $\Delta_o$ is the tunneling matrix element.
The values of these parameters are assumed to vary from TLS to TLS according to the probability distribution:
\begin{equation}
\label{eq:eps-delta-dist}
P(\Delta, \Delta_o)=\frac{\overline{P}}{\Delta_o}
\end{equation}
with \(0<\Delta <\Delta_{\text{max}}\) and \(\Delta_{o,\text{min}}<\Delta_o<\Delta_{o,\text{max}}\). ${\overline{P}}$ is the constant density of states of tunneling entities.
The energy eigenvalues of a given TLS are
\be
\pm\frac{1}{2}E=\pm\frac{1}{2}\sqrt{\Delta^2+\Delta_o^2}.
\ee
Thus the energy splitting between the two levels is $E$.

A TLS with an electric dipole moment $\vec{p}$ interacts with an
electric field $\vec{E}$. The elastic dipole moment of the TLS interacts
with the strain field. Technically, the elastic dipole moment and 
the elastic strain field are tensors. However, we can use a 
simplified Hamiltonian to describe these interactions: 
\begin{equation}
H_{int}= \vec{p}\cdot\vec{E}\sigma_z + \gamma\epsilon s\sigma_z,
\label{Eq:H_TLSFields}
\end{equation}

where $s$ corresponds to half the difference in the elastic dipole moment
between the two wells of the double well potential \cite{Carruzzo2020},
and $\gamma$ is the magnitude of the interaction between 
the TLS and the scalar elastic strain field $\epsilon$.

Experimental and theoretical work indicate
that TLS interact with each other via the elastic strain field 
\cite{Yu1988,Salvino1994,Carruzzo1994}. If the 
TLS have electric dipole moments, then they can also interact 
with one another via electric
fields. However, for reasonable values of the electric dipole moments,
the electric dipole-dipole interaction is about an order of magnitude smaller 
than the elastic dipole-dipole interactions \cite{Yu1985}. 
To demonstrate this, let us estimate the couplings for electric and elastic dipole-dipole interactions. By coupling, we are referring to the
prefactor of the $1/r^3$ dipole term in the Hamiltonian. As we describe below,
the effective interaction between two TLS exchanging phonons goes as $g/r^3$
where the coupling $g\sim\gamma^2/\rho v^2$ where $\rho$ is the density of 
the material, $v$ is the speed of sound, and $r$ is the separation 
between the two TLS \cite{Joffrin1975,Yu1988}. A typical order of magnitude
value for $\gamma$ in glasses is 1 eV \cite{Berret1988,Yu1987}. 
Using the values from SiO$_2$ for
$\rho=2.2$ g/cm$^3$ and $v=4.2\times 10^5$ cm/s, we find $g\sim 5\times 10^4$
K$-\AA^3$ \cite{Yu1988}. (The value for $v$ is a weighted average: 
$3v^{-2}=v_{\ell}^{-2}+2v_{t}^{-2}$ where $v_{\ell}$ is the longitudinal
speed of sound and $v_t$ is the transverse speed of sound.) 
For an electric dipole-dipole interaction, the coupling
is $p^2/(4\pi\epsilon_o)$ where $\epsilon_o$ is the permittivity of the
vacuum. In Suprasil W, a particularly pure form of fused silica, 
the intrinsic dipole moment of TLS is 0.6 D \cite{Golding1979}
while the electric dipole
moment of OH$^-$ impurities in Suprasil I have $p=3.7$ D \cite{Golding1979}.
If we take $p\sim 1$ D, then the electric dipole-dipole coupling is
$7\times 10^3$ K$-\AA^3$. Another example is 
(KBr)$_{1-x}$(KCN)$_{x}$ where the elastic coupling between two CN$^{-}$ ions
is approximately $8\times 10^3$ K $\AA^{3}$ compared to 
$p^2\sim 7\times 10^2$ K $\AA^{3}$, where the CN$^{-}$ dipole moment
is $p\sim 0.3$ D \cite{Yu1985}. Thus we see that the elastic interaction between
TLS is typically about an order of magnitude larger than the electric 
dipole-dipole interaction. 

As mentioned above,
by integrating out the strain field, one can write the interacting Hamiltonian 
as \cite{Joffrin1975,Carruzzo2020}:
\be
H={1\over 2}\sum_{i\not= j}\sigma_i^z\Lambda_{ij}\sigma_j^z.
\ee
Here $\Lambda_{ij}$ is given (in simplified form) by: 
\be
\Lambda_{ij}=\left(\frac{\gamma^2}{\rho v^2}\right)
\frac{s_i s_j}{r_{ij}^3},
\ee
where $r_{ij}$  is the distance between TLS $i$ and $j$ 
(see \cite{Joffrin1975} for the full expressions). We can rewrite the 
Hamiltonian in terms of local fields produced by the neighboring TLS:
\be
H={1\over 2}\sum_{i}h_i\sigma_i^z,
\ee
where the local field operators $h_i$ are given by
\be
h_i=\sum_{j\neq i}\Lambda_{ij}\sigma_j^z.
\ee
Fluctuations in the neighboring TLS produce 
fluctuations the local field and hence in the energy splitting of the
$i$th TLS. This leads to relaxation times that vary in time since the
TLS relaxation rate is given by \cite{Hunklinger1986,Phillips1987}:
\be
\tau^{-1}=\frac{\gamma^2}{\rho}\left[\frac{3}{v^5}\right]
\frac{E^3}{2\pi\hbar^4}\left[\frac{\Delta_o}{E}\right]^2
\coth\left(\frac{\beta E}{2}\right),
\label{eq:TLSrelaxation}
\ee 
where $\beta=1/k_B T$ is the inverse temperature.
As a result, a large number of fluctators with a broad distribution of 
relaxation times can produce 1/f noise 
\cite{Kogan1996,Faoro2006,Constantin2009}. However, a quantum dot will
not be equally affected by a large number of fluctuators. Rather, the
TLS nearest to the QD will have the greatest influence on its charge 
noise spectrum. So we will explore
a different mechanism to explain 1/f charge noise.

Furthermore, it is the electric dipole moments, not the elastic dipole moments,
that couple to quantum dots and give rise to charge noise. We assume that
when a TLS tunnels from one well to the other, both its elastic and electric
dipole moments flip. This results in fluctuations in the elecric 
potential $V(t)$ seen by the QDs. Since there is a metal ground 
plane, both the electric dipole moments $\vec{p}$ and their image charges 
will contribute to $V(t)$ which is given by
\begin{equation}
V_{\zeta}(t)=\sum_{\eta=1}^2 \sum_{i=1}^{N} \frac{\vec{p}_{\eta,i}(t) \cdot \hat{R}_{\eta,i,\zeta}}{R_{\eta,i,\zeta}^2},
\label{eq:Vpotential}
\end{equation}
where $\zeta=1,2$ denotes the QD, and
the sum over $i$ corresponds to summing over TLS in the oxide layer
(denoted by $\eta=1$) 
and their images (denoted by $\eta=2$). $R_{\eta,i,\zeta}$ is the 
distance between the electric dipole $\vec{p}_{\eta,i}$  and the $\zeta$th
quantum dot, and ${\hat R}_{\eta,i,\zeta}$ is a unit vector 
that points from the dipole to the $\zeta$th quantum dot.

We represent the TLS electric dipoles by a 2D square lattice of 
dipoles that are randomly oriented. We set the magnitudes of each dipole
moment to unity. The mirror images of these electric dipoles in the ground
plane form a second 2D square lattice. The dynamics of the TLS is
governed by their elastic interactions which we model very simply with a 2D
nearest neighbor Ising spin glass on a square lattice. Each Ising spin
represents a TLS elastic dipole. When the Ising spin (elastic dipole) flips,
its corresponding electric dipole moment, as well as its image electric dipole 
moment, flips 180$^{\circ}$. In some sense the Ising spin is an avatar or proxy
for the electric dipole moment. 

To understand why 
the elastic and electric dipole moments of a given TLS flip together,
consider the following simple picture. Suppose the TLS consists of 
a charged atom that hops between the two minima of a double well potential. 
When the atom hops, its elastic dipole flips. Since the atom 
is charged, when it hops, its electric dipole moment also flips. 
In fact, a TLS flip could involve a more complicated rearrangement of a 
collection of atoms and charges. The key point is that the atoms can be 
associated with charges. If the atoms rearrange, then the charges rearrange. 
So the elastic and electric dipoles flip together.

The orientations of the electric dipole and elastic dipole are likely 
correlated but there is no reason to favor any particular correlation. 
(Technically, an elastic dipole moment is a tensor while an electric dipole 
moment is a vector, so it is not clear how to compare their orientations.) 
If we think again of our single atom hopping between two positions, 
the electric dipole moment could lie along the line connecting those 
two positions. But a more complicated rearrangement of several atoms could have an electric dipole moment that reorients in some arbitrary direction. 
In any event, reorientation of the elastic dipole moments lead to 
flips of electric dipoles and their images, which
will change the electric potential $V_i(t)$ seen by the $i$th quantum dot 
as given by Eq. (\ref{eq:Vpotential}). 
The time series of $V_i(t)$ is 
Fourier transformed to produce the charge noise power spectrum.

For the purposes of our simulation, it does not matter whether the transitions between the two states of a TLS occur via tunneling or thermal activation because we assume there is no coherence between successive transitions. Thus we represent a TLS elastic dipole by an Ising spin. Since TLS occur at random locations, their interactions are random. So we will use a nearest-neighbor Ising spin glass to model the interacting TLS bath. 

\subsection{Simulation Details}

We performed Monte Carlo simulations of a 2D Ising spin glass on a square
16 $\times$ 16 lattice with periodic boundary conditions. Each Ising spin
corresponds to a TLS with an elastic dipole moment.
The Hamiltonian is given by
\begin{equation}
\label{e_Hamiltonian-Ising}
H=-\sum_{\langle i,j \rangle} J_{ij} S_i S_j,
\end{equation}
where $S_i$ and $S_j$ are Ising spins with values $\pm 1$ on nearest neighbor sites i and j, respectively.  $J_{ij}$ is the nearest neighbor coupling. It represents the elastic coupling between TLS.  We use a spin glass distribution of exchange couplings chosen from a normal distribution centered at $J_{ij}=0$ with a standard deviation $\sigma_{J}=1$. The temperature is in units of $\sigma_{J}$. A positive value for $J_{ij}$ indicates a ferromagnetic interaction.

We can rewrite the Hamiltonian in terms of local fields $h_i$:
\begin{equation}
H=-\frac{1}{2}\sum_i h_i S_i,
\label{e_local-fieldHamiltonian}
\end{equation}
where the local field of $S_i$ is produced by the neighboring spins:
\begin{equation}
h_i=\sum_j J_{ij}S_j.
\label{e_local-field}
\end{equation}

The Ising spins are initialized at a high temperature with 
randomly oriented spins. A spin is allowed to reorient itself according to the Metropolis algorithm~\cite{Metropolis1953}.  In this algorithm, a trial move consists of first choosing a spin on the lattice at random.  For a given temperature $T$, the initial energy $E_i=h_iS_i$ of this site is calculated using Eq. (\ref{e_local-field}) from the local field produced by its nearest neighbors.  The orientation of the spin is reversed in a trial move, and the final energy $E_f$ of this site with the new spin orientation is calculated.  If the final energy is less than the initial energy, then the spin flip is accepted.  However, if the final energy is greater than the initial energy, then the Boltzmann factor $\exp[-(E_f-E_i)/(k_B T)]$ is calculated.  If a random number generated from a uniform distribution between 0 and 1 is less than this Boltzmann factor, then the new orientation is accepted.  This process continues for the remaining sites within the lattice until all the spins in the lattice have been given an opportunity to flip.  The time it takes for one sweep through the lattice is one Monte Carlo step (MCS).

The Ising spin glass is cooled from its initial infinite temperature spin configuration at $T=10$ to $T=0.6$. At each temperature, after an initial equilibration time of $10^5$ MCS, we check to see if the system is equilibrated using the method of Bhatt and Young \cite{Bhatt1988}. Details of the equilibration are given in Appendix~\ref{a_equilibration}.

Each TLS has an electric dipole moment. These electric dipoles are initialized with random orientations. Since the electric dipole-dipole interaction is an order of magnitude smaller than the elastic dipole-dipole interaction, it is the elastic interactions between TLS that governs their dynamics. We represent these dynamics by the Ising spin glass described above. Thus, each electric dipole has a corresponding Ising spin. If the Ising spin flips, then the corresponding electric dipole and its image dipole flip 180$^o$. We consider two
cases of electric dipole orientations. In both cases, the magnitude
of the dipole moment is unity. In the first case, the dipoles are
randomly oriented with random x, y, and z components. To better understand
the contribution of each component, we also consider a second case where the
dipoles all lie parallel to one another along the x, y, or z axis, though
they are randomly oriented in the positive or negative direction.

\begin{figure}
\centering
\includegraphics[width=0.9 \linewidth]{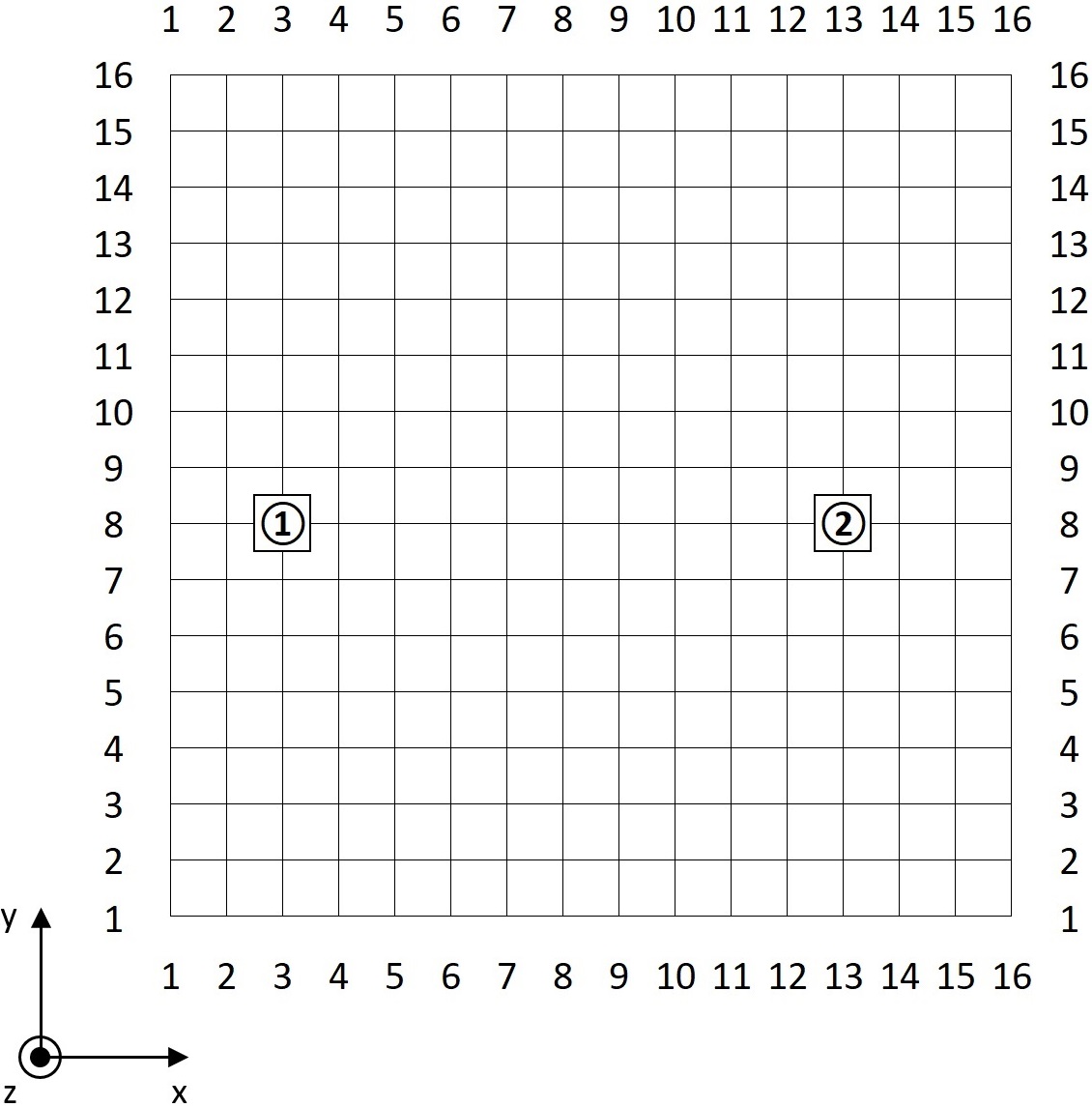}
\caption{16 x 16 lattice with an electric dipole occupying every site.  The circled 1 and 2 represent quantum dots that are 10 lattice spacings apart in a plane that is three lattice spacings beneath the main lattice.}
\label{f-lattice}
\end{figure}

Let us take a moment to describe the device geometry that we have in mind.
TLS that have both an electric and an elastic dipole moment are located 
on every site of a 16 $\times$ 16 square lattice. 
We envision these TLS to reside in the oxide layer of a Si/Si-Ge 
heterostructure with a quantum well.
Since TLS are typically about 10 nm apart 
(see Appendix \ref{appendixTLS}), 
we take the lattice spacing to
be 10 nm; this sets the length scale. A second 16 $\times$ 16 square lattice
is located half a lattice spacing above the first lattice and is populated 
by the image dipoles that result from the ground plane formed by metal gates.
We consider two QDs about 100 nm apart located in a quantum well
about 30-50 nm below the oxide layer. So in our simulation two QDs 
plane are separated by 10 lattice spacings in a plane located 3 lattice
spacings below the first electric dipole lattice. This is illustrated in
Figure \ref{f-lattice}. To see how the correlations of the charge noise depend
on the distance between QDs, we also considered cases where the QDs 
were separated by 3, 5, and 7 lattice spacings in the plane representing the
oxide layer. In all three cases, the y-coordinate of 
the QDs is the same as in Figure \ref{f-lattice} and the QDs were 
placed symmetrically on either side of the center of the lattice.

\subsection{Noise Power, Exponents, and Correlation Functions}

The fluctuating electric dipole moments produce a
time dependent electric potential $V_i(t)$ at the $i$th QD given by 
Eq. (\ref{eq:Vpotential}). After equilibration of the Ising spin glass,
the potential is recorded at every Monte Carlo step, resulting in a time
series that can be Fourier transformed to obtain the charge noise 
power spectrum. 
The deviation from the average is $\delta V_i(t) = V_i(t) - \overline{V_i(t)}$.  The noise power spectral density can be determined from the Fourier transform of the autocorrelation function $C_i(\tau)=\int_{-\infty}^{\infty} \delta V_i(t) \delta V_i(t+\tau)dt$. For a time series of length $t_{\text{total}}$: ${S(f)=\frac{1}{t_{\text{total}}}\int_{-\infty}^{\infty} C_i(\tau) e^{-2 \pi i f \tau} d\tau}$. 
A useful method for computing the power spectrum is the periodogram estimate~\cite{Press1992}:
\begin{equation}
S_i(f)= \frac{1}{t_{\text{total}}} \lvert \delta V_i(f) \rvert^2,
\end{equation}
where $t_{\text{total}}$ is the length of the potential time series, and $\delta V(f)$ is the Fourier transform of the electric potential time series.  The Fourier transform is computed using the C subroutine library FFTW~\cite{Frigo2005}.  At a given temperature, the time series $V_i(t)$ is split into ten segments of equal length.  The power spectrum is found for each segment and is averaged over these segments to give a smoother power spectrum.  The power spectrum is normalized so that
\begin{equation}
\label{e_total-power}
P_{i,\text{total}}=\frac{2}{t_{\text{total}}} \sum_{f=0}^{f_{\text{max}}} S_i(f) = \sigma^2_{V_i},
\end{equation}
where $P_{\text{total}}$ is the total noise power and $\sigma^2_{V_i}$ is the variance in the elecctric potential of the $i$th QD. To determine the noise exponent $\alpha$, the function $A^2/f^\alpha$ is fit to power spectra (averaged over 10 segments) in the vicinity of a fixed frequency because experiments typically determine noise exponents at 1 Hz. 

To calculate the correlation between the fluctuating dipole potentials $V_1(t)$ and $V_2(t)$, we use the Pearson correlation coefficient that is given by:  
\begin{equation}
\rho_{V_1(t),V_2(t)} = \frac{\langle \delta V_1(t) \delta V_2(t) \rangle_t}{\sigma_{V_1(t)} \sigma_{V_2(t)}},
\label{e-correlation}
\end{equation}
where $\delta V_1(t)$ and $\delta V_2(t)$ are the fluctuations in the potentials about their average.  $\sigma_{V_1(t)}$ and $\sigma_{V_2(t)}$ are the standard deviations of the potentials. To calculate the correlations, the time series $V_1(t)$ and $V_2(t)$ are divided into ten blocks of equal size.  The correlation is calculated for each of the ten blocks using Eq. (\ref{e-correlation}), and the standard deviation is calculated from the ten correlations.  The correlations and standard deviations are then averaged over 200 independent runs.

We also calculate the noise correlation between the two QDs as a function of
frequency as was done in \cite{Yoneda2023}. We use:
\begin{equation}
\label{e-correlationf}
\rho_{V_1(t),V_2(t)}(f) = \lfrac{\langle \delta \tilde{V}_1(f) \delta \tilde{V}^*_2(f) \rangle_{\text{blocks,runs}}}{\sqrt{\left\langle S_{V_1}(f) \rangle_{\text{blocks,runs}} \langle S_{V_2}(f) \right\rangle_{\text{blocks,runs}} }},
\end{equation}
where $\delta{\tilde{V}_1(f)}$ and $\delta{\tilde{V}_2(f)}$ are the Fourier 
transforms of the fluctuations in the potentials about their average.  
$S_{V_1}(f)$ and $S_{V_2}(f)$ are the noise power of the dipole potentials at 
quantum dots 1 and 2.  To calculate the numerator of 
Eq. (\ref{e-correlationf}), the time series $V_1(t)$ and $V_2(t)$ are 
divided into 100 blocks of equal size.  For each block, the Fourier 
transforms of the fluctuations are calculated.  The product 
$\delta \tilde{V}_1(f) \delta \tilde{V}^*_2(f)$ is averaged over the 
100 blocks and 200 runs to give $\langle \delta \tilde{V}_1(f) \delta \tilde{V}^*_2(f) \rangle_{\text{blocks,runs}}$.  To calculate the denominator of 
Eq. (\ref{e-correlationf}), the power spectra $S_1(f)$ and 
$S_2(f)$ are calculated for each block.  Then $S_1(f)$ and
$S_2(f)$ are each averaged over the 100 blocks 
and 200 runs to give $\langle S_{V_1}(f) \rangle_{\text{blocks,runs}}$ and 
$\langle S_{V_2}(f) \rangle_{\text{blocks,runs}}$.  The denominator of the 
correlation is then calculated and used in Eq. (\ref{e-correlationf}).
Since the frequency-dependent correlation function in 
Eq. (\ref{e-correlationf}) is a complex number, we plot its magnitude and
phase as a function of frequency.

\section{Results}

\begin{figure}
\centering
\includegraphics[width=1.0 \linewidth]{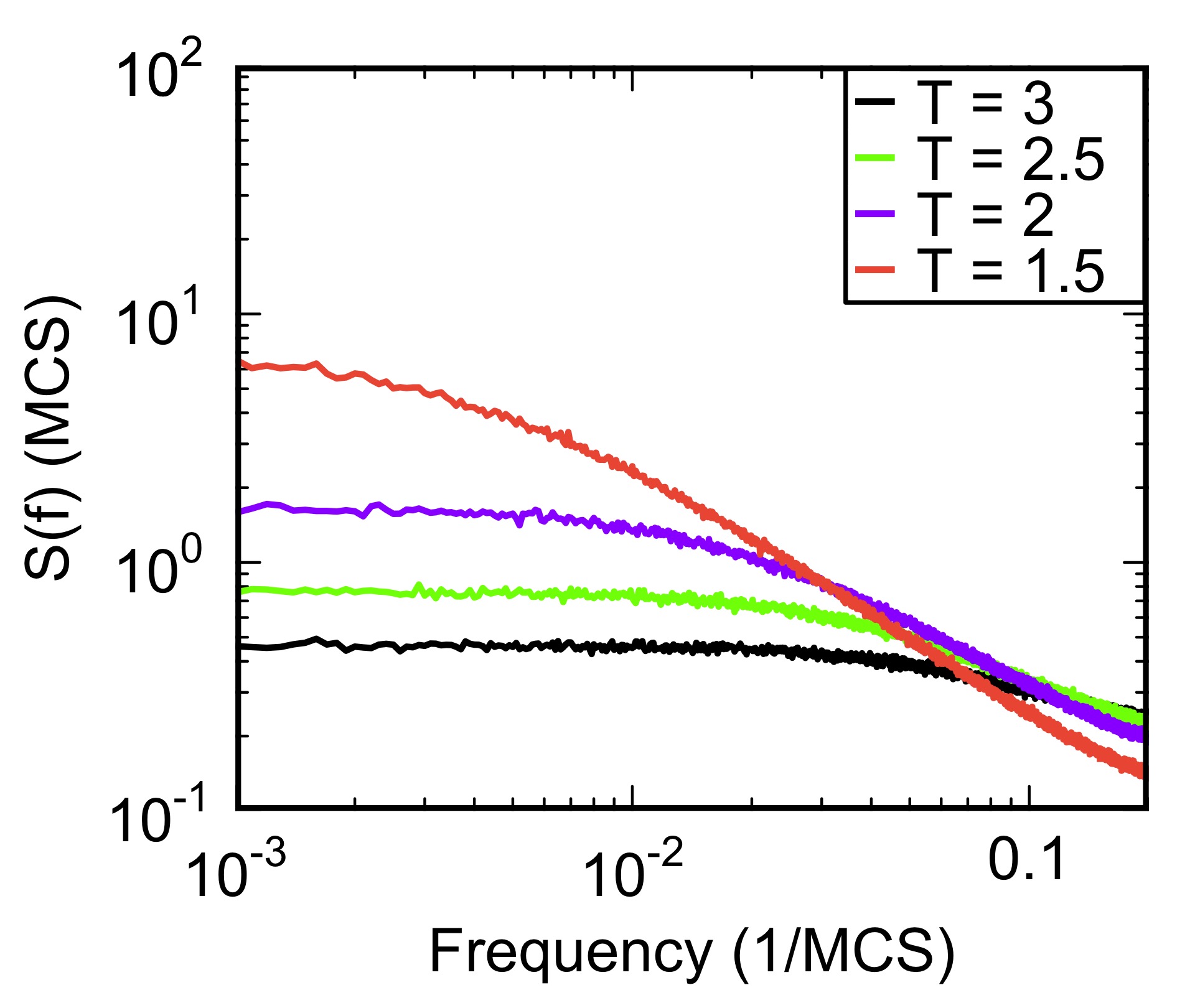}
\caption{Plots of the electric dipole potential noise power at QD 1 vs. 
frequency for $1.5 \leq T \leq 3$ averaged over 200 runs. 
The noise is produced by fluctuating electric dipoles with random 
orientations. The QDs are 10 lattice spacings apart.}
\label{f-power2}
\end{figure}

\begin{figure}
\centering
\includegraphics[width=1.0 \linewidth]{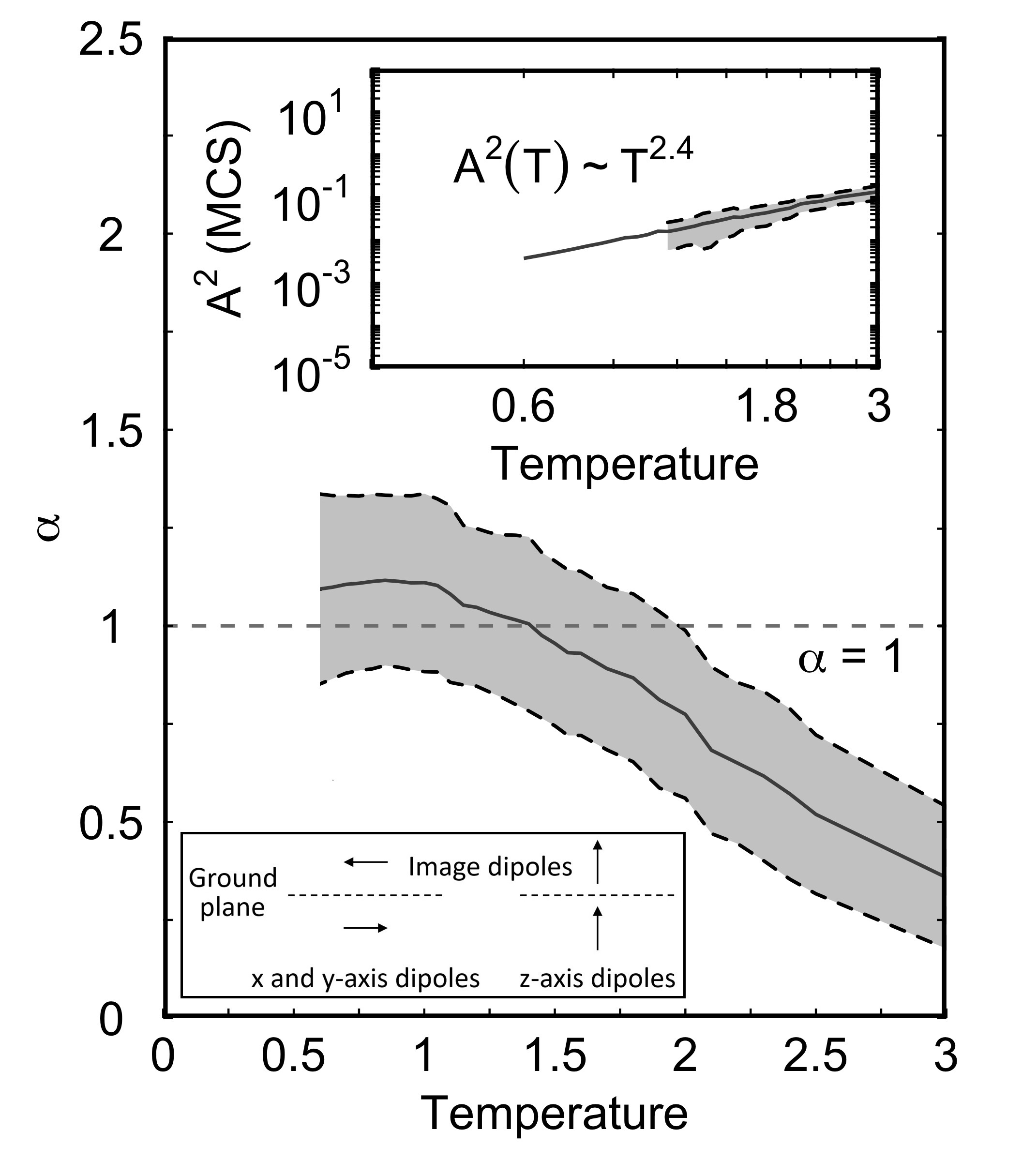}
\caption{Plot of the temperature dependence of the noise exponent, $\alpha (T)$, of the electric potential noise power spectra at QD 1 resulting from fluctuating electric dipoles with random orientations for $0.6 \leq T \leq 3$.  The solid line in the middle is the average noise exponent over 200 runs. The top and bottom dashed lines are one standard deviation above and below the mean. Upper inset: Log-log plot of the temperature dependence of the electric potential noise power amplitude, $A^2(T)$, at QD 1 resulting from fluctuating electric dipoles with random orientations for $0.6 \leq T \leq 3$. The central solid line is the amplitude averaged over 200 runs. The top and bottom dashed lines of the shaded region are one standard deviation above and below the mean amplitude. The lack of shading at low temperatures is due to the standard deviation being greater than the mean which cannot be shown on a logarithmic scale. The QDs are 10 lattice spacings apart in the main plot and in the upper inset. Lower inset: Orientation of dipoles 
and their dipole images.}
\label{f-Aa2}
\end{figure}

Figure \ref{f-power2} shows the noise power at QD 1 at various temperatures for randomly oriented dipoles averaged over 200 runs. To determine the amplitude $A^2(T)$ and the noise exponent $\alpha(T)$, the function $A^2(T)/f^{\alpha(T)}$ is fit to the region of the power spectra that is linear on a log-log plot. The noise exponent $\alpha (T)$ as a function of temperature is shown in Fig.~\ref{f-Aa2}. Experimentally, in the temperature range from 50 mK to 1 K, the noise exponents within one standard deviation of the mean at 1 Hz ranged from about 0.65 to 1.3 ~\cite{Connors2019,Jock2022}. In our simulations the noise exponents increase from zero at high temperatures to approximately 1.15 at low temperatures which is roughly consistent with experiment.

The noise amplitude $A^2(T)$ shown in the inset of Fig.~\ref{f-Aa2} goes as $T^{2.4}$. For an ensemble of fluctuators with thermally activated switching rates and a flat distribution of activation energies, one would expect the noise amplitude to increase linearly with temperature \cite{Petit2020}. However, as we mentioned in the introduction, experiments find the temperature dependence of the noise at 1 Hz deviates from linearity \cite{Connors2019,Connors2019e,Petit2018,Spence2023}. In one study, the temperature dependence increasingly deviates from linearity with increasing thickness of the gate oxide layer \cite{Connors2019,Connors2019e} and, in another study,
was found to be quadratic \cite{Petit2018}. So the 
lack of a linear temperature dependence in our simulation results is
qualitatively consistent with experiment, implying that the distribution of barrier heights is not uniform but the reason why is not clear.
We will consider this further in the discussion section.

\begin{figure}
\centering
\includegraphics[width=0.97 \linewidth]{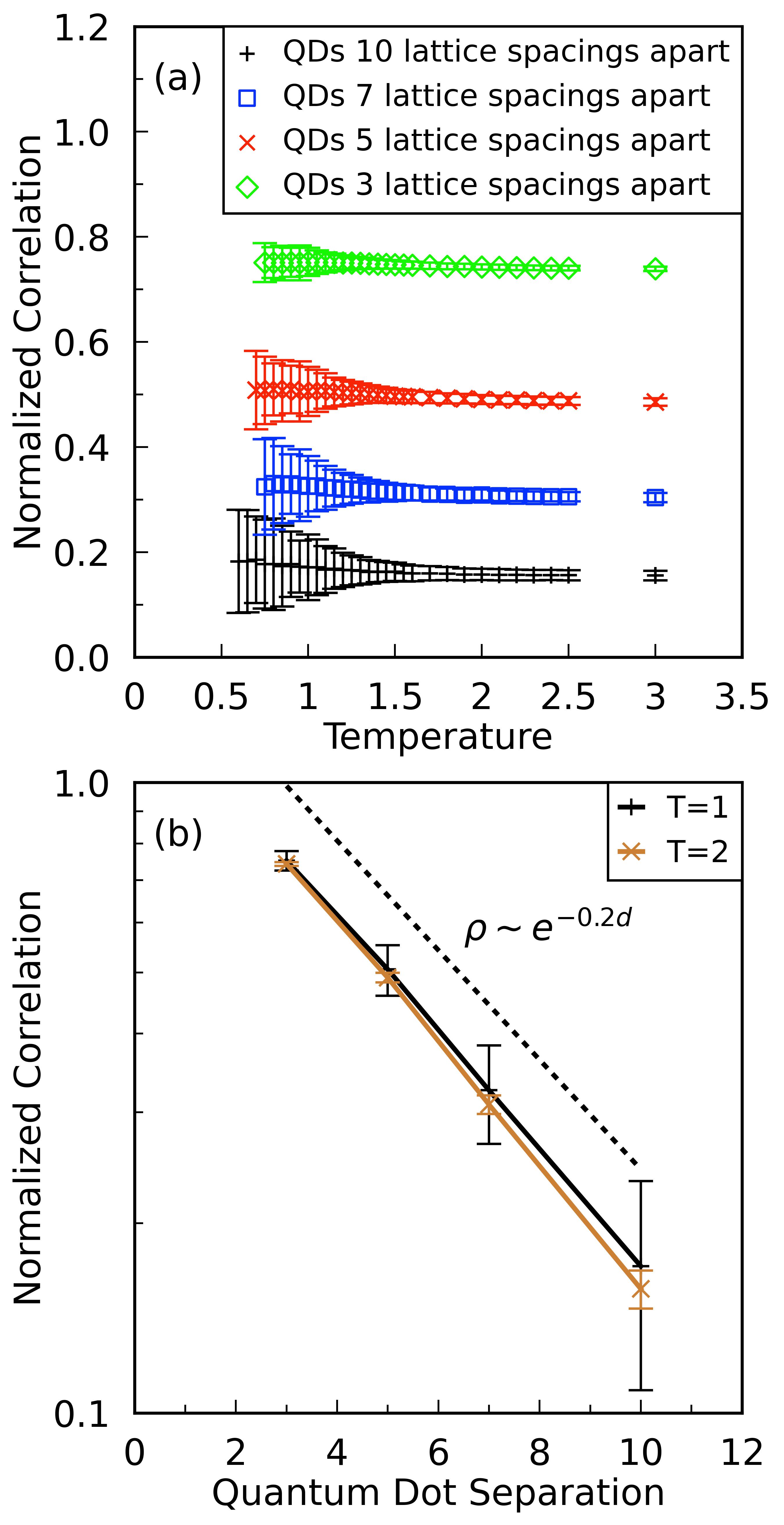}
\caption{(a) Plot of electric potential noise correlations between QDs 1 
and 2 vs. temperature for $0.8 \leq T \leq 3$ 
resulting from fluctuating electric dipoles with random orientations. 
QD separations of 3, 5, 7, and 10 lattice spacings are shown.
(b) Lin-log plot of the charge noise correlations between QDs as a function of 
QD separation at $T=1$ and $T=2$. The noise correlations decrease 
exponentially with increasing QD separation $d$. 
The error bars shown in (a) and (b) correspond to the standard deviation that is calculated as described in the text.
}
\label{f-correlation2}
\end{figure}

If one considers the contributions of the different components of the dipoles to the charge noise, one realizes that the z-component dominates because the z-component of the image dipole in the ground plane (that lies perpendicular to the z-axis) points in the same direction as the z-component of the original dipole (see lower inset in Figure \ref{f-Aa2}). However, the x and y components of the image dipole point in the opposite direction of the x and y components, respectively, of the original dipole (see lower inset in Figure \ref{f-Aa2}). To confirm this, we have performed simulations for dipoles that lie along either the x, y, or z axis (see Appendix~\ref{a_xyz-dipoles} for details of the simulations). The results are shown in Figures ~\ref{f-panel1} and \ref{f-panel2} in Appendix~\ref{a_xyz-dipoles}. One can see that the noise produced by the dipoles along the z-axis is about two orders of magnitude larger than that associated with dipoles along the x or y axes. Comparing the plots in Fig. \ref{f-panel1} with the corresponding plots (Figs. \ref{f-power2}, \ref{f-Aa2}, and \ref{f-correlation2}) for randomly oriented dipoles, we see the marked similarity between the results for the dipoles along the z-axis and the randomly oriented dipoles. However, the noise power and noise amplitudes resulting from electric dipoles aligned along the z-axis are three times larger than that of the randomly oriented dipoles because the z-axis dipoles have unit length, while the z components of the randomly-oriented dipoles have a typical length of $1/\sqrt{3}$.

Using Eq. (\ref{e-correlation}), we calculated
the correlation in the noise for various separations between QD 1 and QD 2 as shown in Fig.~\ref{f-correlation2}. The correlation decreases exponentially with increasing distance between the QDs. At a separation of 10 lattice spacings corresponding to about 100 nm, we can see from Fig.~\ref{f-correlation2} that there is some correlation between QDs 1 and 2 which is consistent with experiment. As we mentioned earlier, recent measurements find that there are correlations in the charge noise between quantum dots that are about 100 nm apart \cite{Yoneda2023} and 200 nm apart \cite{RojasArias2023}. Furthermore, Ref. \cite{RojasArias2023} found that charge noise correlations decayed exponentially up to a few hundred nm and then decayed as a power law $\sim d^{-a}$ where $d$ is the QD separation and the exponent $a$ varied between 4 and 5.

In Fig.~\ref{f-correlation2}(a), the error bars get larger as the temperature decreases because the spins (TLS) fluctuate more slowly at lower temperatures, leading to bigger variations between time slices in the time series data and hence, to bigger error bars. These are the same slow fluctuations that produce low frequency noise.

\begin{figure}
\centering
\includegraphics[width=1.0 \linewidth]{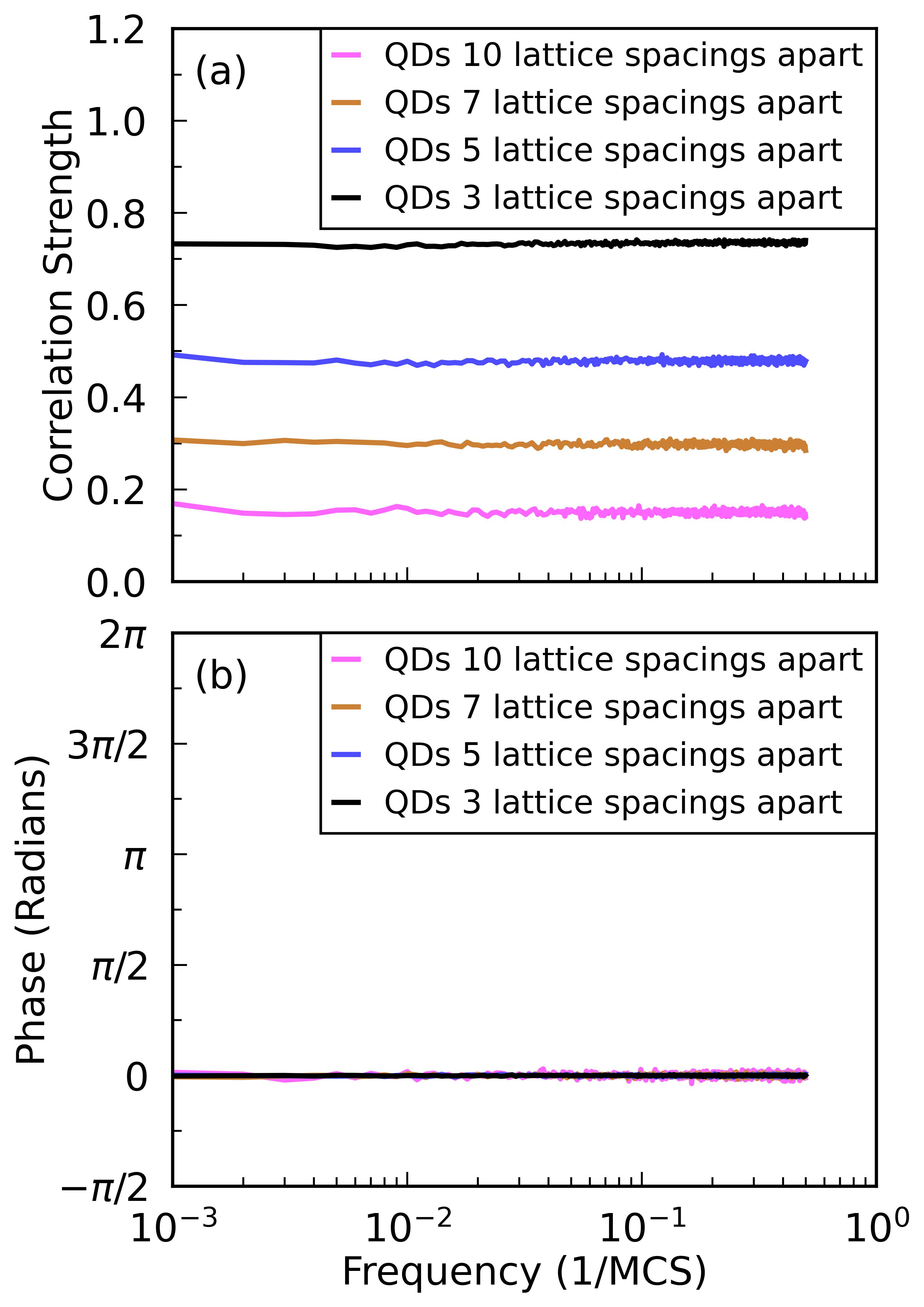}
\caption{Plots of the (a) magnitude and (b) phase of electric potential noise 
correlations between QDs 1 and 2 vs. frequency at $T=1$ for QD separations
of 3, 5, 7, and 10 lattice spacings. 
The noise correlations decrease with increasing distance.
The noise arises from fluctuating electric dipoles with random orientations.
}
\label{fig:correlationMagPhase}
\end{figure}

We used Eq. (\ref{e-correlationf}) to calculate the frequency
dependence of the magnitude and phase of correlations in the
noise produced by randomly oriented fluctuating electric dipoles.
These results are shown in Figure \ref{fig:correlationMagPhase}
for QDs separated by distances of 3, 5, 7, and 10 lattice spacings.

\begin{figure}
\centering
\includegraphics[width=0.9 \linewidth]
{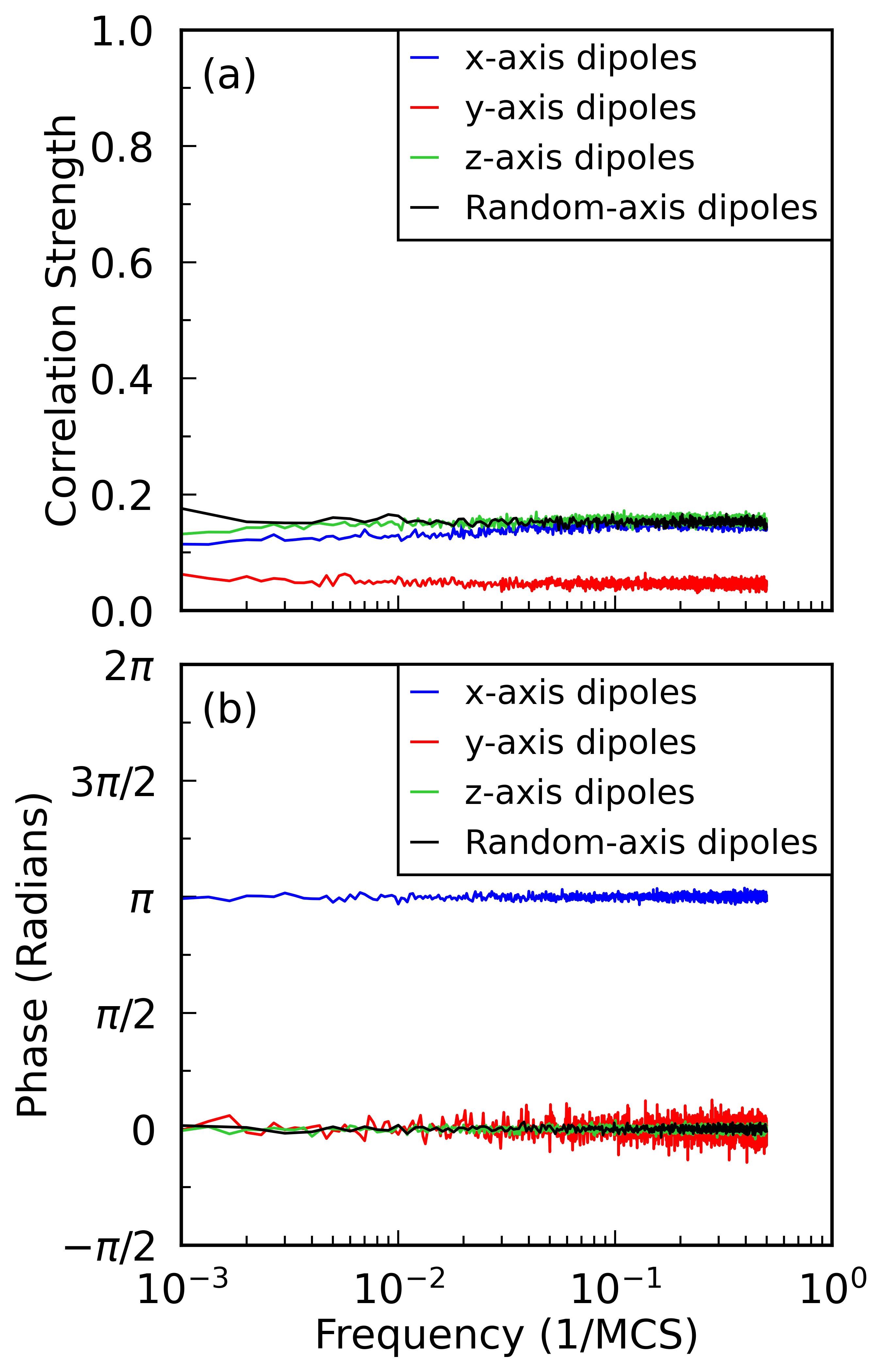}
\caption{Plots of the (a) magnitude and (b) phase of electric potential
noise correlations between QDs 1 and 2 that are 10 lattice spacings apart
as a function of frequency 
for fluctuating electric dipoles with fixed (x, y, and z) and random 
orientations at $T = 1$. Plots are the result of averaging over 200 runs.}
\label{f-correlationf}
\end{figure}

To better understand these correlations in the noise, we
considered the more realistic case of two QDs separated by 10 lattice spacings.
We used Eq. (\ref{e-correlationf}) to calculate the frequency 
dependence of the magnitude and phase produced by randomly 
oriented fluctuating electric dipoles as well as dipoles aligned along
various axes (see Figure \ref{f-correlationf}). Fig.
\ref{f-correlationf}(a) shows that there is correlation in the noise
between the two dots as we saw in the Pearson correlation. 
These correlations are due largely to the geometry of the location of the 
dipoles with respect to the QDs. There is no noticeable frequency dependence 
in either the magnitude or the phase of the correlations because
unlike the experimental case \cite{Yoneda2023}, our QDs have no dynamics.

The magnitude of the noise correlation is larger for the x-axis and z-axis dipoles compared to the y-axis dipoles, and comparable to that for randomly
oriented dipoles (see Figs. \ref{f-panel1}(f) and \ref{f-correlationf}(a)).
Even though the numerator of Eq. (\ref{e-correlationf}) is about two orders 
of magnitude larger for dipoles aligned along the z-axis compared to those
aligned along the x-axis, the normalization by the noise amplitude in the
denominator makes the noise correlation of the x and z axis dipoles comparable.
The sizeable contribution of the x-axis dipoles, especially those located at
or near the same y-coordinate as the QDs, is due to the large value of 
($\vec{p}_{\eta,i}(t)\cdot\hat{R}_\eta$), the numerator of 
Eq. (\ref{eq:Vpotential}), coupled with the small value of the denominator
$R_{\eta,i}^2$. ($\vec{p}_{\eta,i}(t)\cdot\hat{R}_\eta$) is large when 
$\vec{p}_{\eta,i}(t)$ is collinear with $\hat{R}_\eta$. On the other hand,
the y-axis dipoles have small values
of ($\vec{p}_{\eta,i}(t) \cdot \hat{R}_{\eta}$) when the y-coordinate of their location
is comparable to that of the QDs because the y-axis dipole is almost
perpendicular to the unit vector $\hat{R}_\eta$ pointing from the dipole to the 
QD. Again z-axis dipoles give large 
contributions because their image charges lie in the same direction as the
z-axis dipoles themselves. 

Figure \ref{f-correlationf}(b) shows that the correlations produced by 
the x-axis dipoles are 180$^{\circ}$ out of phase. This is due to the fact that
since the QDs lie along the x-axis, flips in the x-axis dipoles located 
between the QDs will have the opposite effect on each dot due to the factor 
($\vec{p}_{\eta,i}(t) \cdot \hat{R}_\eta$) in the numerator of 
Eq. (\ref{eq:Vpotential}).

\section{Discussion}
We have proposed a model in which the 1/f charge noise in quantum dots is due to a bath of electric dipole fluctuators that interact with each other primarily
via the elastic strain field. We use a 2D nearest-neighbor Ising spin glass to represent these elastic interactions and to simulate the dynamics of the bath of electric dipole fluctuators in the presence of a ground plane representing metal gates above the oxide layer containing the fluctuators. We find 1/f noise
spectra with a nonlinear temperature dependent amplitude, a 
feature that has also been seen experimentally.
Likewise, the noise correlations between
quantum dots are in qualitative agreement with experiment, with
the correlations decreasing exponentially with increasing QD separation.

Unlike previous theoretical models which focused on the effects on qubits
of charge noise produced by independent charge fluctuators, we focused
on interacting TLS with only placeholders for qubits. Interactions between
TLS cause the dynamics and energy splitting of individual TLS to change
with time, leading to noise spectra that have a range of noise exponents as
has been observed experimentally.

We simulated an Ising spin glass on a 16 x 16 lattice. 
One might wonder whether
our results would change if we had used a larger lattice. We believe that 
1/f noise will still occur for different size lattices as well as 
for randomly placed 
TLS. Going to larger lattice sizes could potentially affect our results in
two ways.  First, a larger number of TLS could affect the QD. This
number would grow roughly as $R^2$, where $R$ is the QD-TLS separation, 
while their influence on the dipolar potential energy noise spectra 
would decrease 
as $R^{-4}$, leading to an overall effect that decreases as $R^{-2}$. Thus
we would expect that the effect of going to larger lattice sizes would quickly
saturate. We have confirmed this lack of size dependence with
Ising spin glass simulations in which a QD was located in the 
center of $L\times L$ lattices where $L$ = 16, 24, and 32 for temperatures 
$T$ between 1 and 10. We found negligible changes in the noise spectra 
for different lattice sizes.

The second way in which going to larger lattice sizes can affect the noise
spectra is the frequency range over which 1/f noise is observed in our
simulations at low temperatures. To understand this, note that 
a QD is so small that it is strongly influenced only by a few nearby TLS.
However, over time, other TLS flip and produce a variety of relaxation times 
in those TLS that strongly influence the QD. The inverse of 
the longest relaxation time in the system corresponds to 
the low frequency knee of the 1/f potential energy noise spectra. 
(The knee is where there is a crossover, as the frequency decreases, 
from 1/f noise to a flat noise spectrum (white noise) at low frequencies.)
The longest relaxation time depends on the correlation length of the
interacting fluctuators. (One can think of the correlation length as
being the size of the biggest cluster of spins that can flip. The bigger
the cluster is, the longer it takes to flip, and hence the longer the
relaxation time.) At high temperatures
(outside the critical region) where the correlation length is shorter than
the lattice size, going to larger lattices will not lead to any significant
change in the noise spectrum \cite{Ogielski1985}.
However, this is not the case at low temperatures in the
critical region, where the temperature is close 
to the phase transition temperature $T_g$ of the spin glass 
(or dipolar glass) \cite{Ogielski1985}. For reference, 
$T_g=0$ for a 2D Ising spin glass \cite{Bhatt1988} 
and for a 3D dipolar glass \cite{Snider2005,Quilliam2007,Jonsson2007}.
At low temperatures in the critical regime, the correlation length exceeds 
the size of the lattice used in simulations. In this case,
going to larger lattice sizes would lead to longer relaxation times
\cite{Ogielski1985} and to lower knee frequencies, i.e., to a
larger frequency range over which there is 1/f noise \cite{Chen2007}.

We found that the noise amplitude $A^2(T)$ goes as $T^{2.4}$ as shown in
the inset of Fig.~\ref{f-Aa2}.  This implies a nonuniform 
distribution of barrier heights \cite{Connors2019,Connors2019e}. We
cannot deduce the distribution of barrier heights from our simulations. 
Ising spins do not have activation barriers per se; they simply flip based
on the energy difference between the initial and final states. The 
distribution of local fields $P(h)$ is analogous to the distribution of
TLS asymmetry energies, but it does not capture the barrier heights of the
energy landscape. In our simulations, the low frequency behavior of charge 
noise involves multiple spin flips and dynamics that explore the TLS energy 
landscape at long time scales. 

As we mentioned earlier, the other mechanism
by which TLS produce slow fluctuations is via tunneling through a barrier,
leading to long relaxation times as given in Eq. (\ref{eq:TLSrelaxation}).
A broad distribution of relaxation rates associated with noninteracting
tunneling TLS produces 1/f
charge noise that increases linearly with temperature, i.e., 
$S(f)\sim T/f$ \cite{Kogan1996,Faoro2006,Constantin2009}.
However, a quantum dot is not equally coupled to a large number
of fluctuators, i.e., it does not equally weight the broad distribution of 
relaxation times. Rather, the TLS (and their images) closest to a QD have 
the greatest influence on its charge noise. The more distant TLS affect 
the QD less, 
though the number of these distant TLS grows as the square of the distance, 
assuming a constant density in the 2D oxide layer. Furthermore,
as we mentioned in the introduction, the experimentally measured 
temperature dependence deviates from linearity
\cite{Connors2019,Connors2019e,Petit2018,Spence2023}. In one
case, the temperature dependence increasingly deviates from linearity with 
increasing thickness of the gate
oxide layer \cite{Connors2019,Connors2019e} and, in another study,
was found to be quadratic \cite{Petit2018}. In our simulations, 
the interactions between the TLS cause their energy splittings and
flipping rates to change, leading to 1/f noise with an amplitude that
increases as $A^2\sim T^{2.4}$. 

In summary, our work highlights the importance of two level fluctuators
and their mutual interactions in producing 1/f charge noise in semiconducting qubits.

\acknowledgments
We thank Sue Coppersmith for bringing this problem to our attention and for helpful discussions. This work was performed in part at the Aspen Center for Physics, which is supported by National Science Foundation grant PHY-2210452.

\appendix
\section{Estimate of Separation Between Two Level Systems}
\label{appendixTLS}

We stated in the text that the typical separation between TLS is 10 nm. In
this appendix, we give the steps leading to this estimate. The basic strategy
is to estimate the density of TLS by integrating the TLS density of states 
$n_o$ derived from specific heat measurements up to 10 K. We choose 10 K, as 
opposed to, say, 50 K, because TLS with energy splittings much greater
than typical experimental temperatures ($\sim$ 100 mK)
will be frozen out in their
ground state and will not flip on experimental time scales. For SiO$_2$,
$n_o=0.842\times 10^{33}$/erg-cm$^3=1.16\times 10^{-4}$/(K-nm$^3$) 
\cite{Yu1987}. Since the TLS density of states is flat (independent of energy)
\cite{Anderson1972, Phillips1972}, we can multiply $n_o$ by 10 K 
to obtain an estimate of the TLS density: 
$\rho_{TLS}\sim n_o\times (10\;{\rm K})\sim 
1.16\times 10^{-3}$/nm$^{3}$. So the distance between neighboring TLS is
approximately 
\begin{equation}
d_{TLS}\sim \left(\rho_{TLS}\right)^{1/3}\sim 10 \;{\rm nm}
\end{equation}
This is comparable to the rule of thumb that the density of TLS is roughly
100 ppm.

We should point out that there are two TLS density of states:
$n_o$ and ${\overline P}$. $n_o$ is obtained from the coefficient of the
temperature in the linear term in the specific heat. (Recall that the 
specific heat of amorphous insulating materials is linear in temperature 
below 1 K.) $n_o$ essentially includes all the TLS,
assuming that all the TLS contribute to the specific heat. 
${\overline P}$ is the subset of TLS that flip fast enough to scatter
phonons and affect
thermal conductivity and ultrasonic attenuation. (Heat transport
in amorphous insulators is due phonons \cite{Zaitlin1975}.) Typically
$n_o$ is about an order of magnitude larger than ${\overline P}$. The
relation between $n_o$ and ${\overline P}$ is given by 
\cite{Black1978,Zimmermann1981}:
\begin{equation}
c(T,t)=\frac{\pi^2}{12}k_BT{\overline P}\ln\left(\frac{4t}{\tau}\right)
\end{equation}
where $c(T,t)$ is the time dependent specific heat and $\tau$ is the
minimum relaxation time for TLS. The specific heat increases
logarithmically with time due to slowly relaxing TLS that have large
tunneling barriers.

We note that some papers \cite{Culcer2013,Shehata2023}
that attribute charge noise
at the quantum dot to a 2D layer of noninteracting TLS use a TLS density 
$\rho\sim 10^{11}$/cm$^2$ which corresponds to a TLS separation of about 30 nm,
while other papers \cite{Kepa2023,KepaCorrelations2023} use 
$\rho\sim 10^{10}$/cm$^2$ corresponding to a TLS separation of 100 nm.
These estimates are based on ${\overline P}$ obtained from thermal relaxation
measurements of vitreous silica at low temperatures ($\sim$ 200 mK)
\cite{Zimmermann1981}. The TLS involved in those experiments
tunnel on the time scale ($\sim 10^3$ s) of the measurements.  
We have used $n_o$ 
because we are interested in TLS interacting with one another over a 
wide range of time scales.

\section{Equilibration}
\label{a_equilibration}
\subsection{Equilibration and Recording Times}
\label{a_times}
The Ising spin glass systems are equilibrated for $10^5 \text{ MCS}$ at $T=10$.  As the system is cooled, the equilibration and recording times are increased if the system is not in equilibrium.  These times as a function of temperature are shown in Table~\ref{t_times}.

\begin{table}[h!]
	\setlength{\tabcolsep}{6pt}
	\renewcommand{\arraystretch}{1.1}
	\centering
	\begin{tabular}{ |m{2.3cm}|m{3.6cm}| } 
		\hline
		 & Equilibration and \\
		Temperature & Recording Times (MCS)  \\
		\hline
		$1 \leq T \leq 10$  & $10^5$ \\
		\hline
		$0.8 \leq T \leq 0.95$  & $3 \times 10^5$ \\
		\hline
		$T = 0.75$  & $10^6$ \\
		\hline
		$0.65 \leq T \leq 0.7$  & $3 \times 10^6$ \\
		\hline
		$T = 0.6 $  & $10^7$ \\
		\hline
	\end{tabular}
	\caption{Equilibration and recording times for the 2D ($16 \times 16$) Ising spin glass with random electric dipole orientations for $0.6 \leq T \leq 10$.}
	\label{t_times}
\end{table}

\subsection{Spin Glass Susceptibility}
\label{a_spinglasssusceptibility}
The test for equilibration follows Bhatt and Young's procedure for the equilibration of Ising spin glasses~\cite{Bhatt1988}.  Two independent replicas of each system with the same exchange couplings are created and run in parallel.  The initial spin configurations for the two replicas are different and random.  For the set of spins \{$\text{s}_i$\} with $N$ lattice sites, the spin autocorrelation function for the replica $n$, after an equilibration time $t_0$, is
\begin{equation}
\label{e-Q}
Q^{(n)}(t)=\frac{1}{N} \sum_{i=1}^N \text{s}_i^{(n)}(t_0) \cdot \text{s}_i^{(n)}(t_0+t),
\end{equation}
where the summation is over all lattice sites.  The spin glass susceptibility for replica $n$ is calculated as the second moment of this overlap and then averaged over 200 different realizations of bonds and anisotropy axes.  This disorder average is denoted by $\left[ \ldots \right]_{\text{av}}$:
\begin{equation}
\label{e-chisg1}
\chi_{\text{SG}}^{(n)}(t)=\frac{1}{N} \left[ \left( \sum_{i=1}^N \text{s}_i^{(n)}(t_0) \cdot \text{s}_i^{(n)}(t_0+t) \right)^2 \right]_{\text{av}}.
\end{equation}

The equilibration time $t_0$ is chosen from the sequence $1,3,10,30,100,300, \ldots{ }$, etc.  The idea is to compare $\text{s}_i^{(n)}(t_0)$ to $\text{s}_i^{(n)}(t_0+t)$ as ${t \to \infty}$ to see whether $\text{s}_i^{(n)}(t_0+t)$ has lost its ``memory" of $\text{s}_i^{(n)}(t_0)$.  In practice, the comparison is done as ${t \to t_0}$.  The spin glass susceptibility in Eq.~(\ref{e-chisg1}) is averaged over a length of time $t_0$:  
\begin{equation}
\label{e-chisg3}
\chi_{\text{SG}}^{(n)}=\frac{1}{N t_0} \left[ \sum_{t=t_0}^{2 t_0-1} \left( \sum_{i=1}^N \text{s}_i^{(n)}(t_0) \cdot \text{s}_i^{(n)}(t_0+t) \right)^2 \right]_{\text{av}}.
\end{equation}

\noindent The summation over $t$ starts at $t_0$ so that the distribution of $Q^{n}(t)$ is Gaussian.  The correlation of the spins at shorter times makes the distribution deviate from a Gaussian.

For small values of $t_0$ and when the system is at low temperatures, there are few spin fluctuations, so $Q^{(n)}(t) \sim 1$ and $\chi_{\text{SG}}^{(n)}(t) \sim N$.  This is in agreement with simulations.

We can also calculate $\chi^{(n)}_{\text{SG}}$ in the high-temperature limit.  We start with two Ising spins $\text{s}_1$ and $\text{s}_2$ that represent $\text{s}_i^{(n)}(t_0)$ and $\text{s}_i^{(n)}(t_0+t)$, respectively, in Eq.~(\ref{e-chisg3}).  The average square of the dot product is calculated as
\begin{equation}
\label{e-ssai}
\begin{split}
\langle \left( \text{s}_1 \cdot \text{s}_2 \right)^2 \rangle_{\text{Ising}} &= \langle \left( \pm s_1 s_2 \right)^2 \rangle \\
&= \langle s_1^2 s_2^2 \rangle \\
&= 1.
\end{split}
\end{equation}

Combining Eq.~(\ref{e-ssai}) with Eq.~(\ref{e-chisg3}), for high temperatures, we get $\chi^{(n)}_{\text{SG}} = 1$ which is seen in simulations.

We then define the average of the two single replica susceptibilities
\begin{equation}
\label{e-chisga}
\overline{\chi}_{\text{SG}}=\frac{1}{2}\left( \chi_{\text{SG}}^{(1)}+ \chi_{\text{SG}}^{(2)} \right)
\end{equation}
\noindent as the two times spin glass susceptibility.

The spin glass susceptibility may also be calculated from the spin overlap of the two different replicas.  The mutual overlap between the spins $\text{s}_i^{(1)}$ and $\text{s}_i^{(2)}$ of the two replicas is 
\begin{equation}
\label{e-Q'}
Q(t)=\frac{1}{N} \sum_{i=1}^N \text{s}_i^{(1)}(t_0+t) \cdot \text{s}_i^{(2)}(t_0+t).
\end{equation}

The spin glass susceptibility is calculated from the spin overlap:
\begin{equation}
\label{e-chisg4}
\chi_{\text{SG}}=\frac{1}{N t_0} \left[ \sum_{t=t_0}^{2 t_0-1} \left( \sum_{i=1}^N \text{s}_i^{(1)}(t_0+t) \cdot \text{s}_i^{(2)}(t_0+t) \right)^2 \right]_{\text{av}}.
\end{equation}

\begin{figure}
\centering
\includegraphics[width=\linewidth]{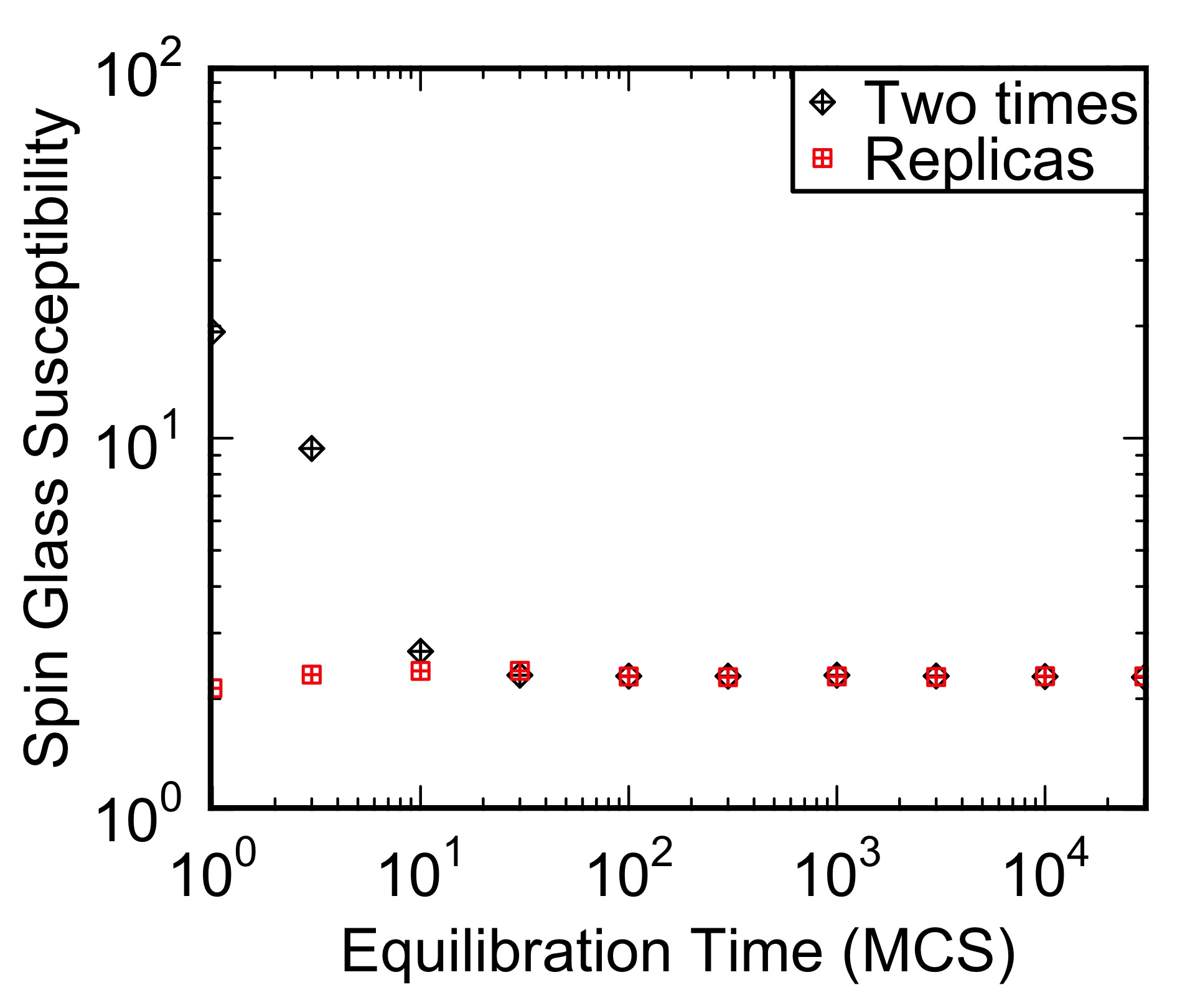}
\caption{Two times and replica susceptibility for the 2D ($16 \times 16$) Ising spin glass for $T=2$ averaged over 200 runs.  For sufficiently long equilibration times, the two susceptibilities agree, and the system is in equilibrium.}
\label{f-spin-glass-susceptibility}
\end{figure}

For all temperatures, as the equilibration time is approached, the spin glass susceptibilities converge; the two times susceptibility $\overline{\chi}_{\text{SG}}$ (Eq.~(\ref{e-chisga})) approaches the true susceptibility from above and the replica susceptibility $\chi_{\text{SG}}$ (Eq.~(\ref{e-chisg4})) from below.  This is shown in Fig.~\ref{f-spin-glass-susceptibility} for the Ising spin glass at $T=2$.

After sufficiently long equilibration times, $\chi_{\text{SG}}$ and $\overline{\chi}_{\text{SG}}$ agree.  We define the system to be equilibrated if
\begin{equation}
\Delta \chi_{SG} =  \lfrac{|\chi_{\text{SG}}-\overline{\chi}_{\text{SG}}|}{\frac{1}{2} \left( \chi_{\text{SG}}+\overline{\chi}_{\text{SG}} \right) } 
\end{equation}
is less than 5\% for three consecutive times in the $t_0$ sequence; then we declare it equilibrated at the fourth time.  For example, if the last three equilibration times are $t_1=3 \times 10^3$, $t_2=10^4$, $t_3=3 \times 10^4$, then the equilibration time $t_4=10^5$.  At each temperature, the initial equilibration time is $10^5$ MCS, and it is increased if the system is not equilibrated.

\subsection{Specific Heat and Susceptibility}
As a check of this equilibration method, we calculate the block-averaged specific heat and magnetic susceptibility in a similar method to Yu and Carruzzo~\cite{Yu2004}.  We use the form of specific heat
\begin{equation}
\label{e_cv}
\begin{split}
C_V &= \frac{1}{k_B T^2} \left( \langle E^2 \rangle - \langle E \rangle^2 \right) \\
&= \frac{N_{\text{sites}}^2}{k_B T^2} \left( \langle e^2 \rangle - \langle e \rangle ^2 \right),
\end{split}
\end{equation}
where $k_B=1$, $T$ is temperature, $E$ is the total energy of the lattice, $N_{\text{sites}}$ is the total number of sites, and $e$ is the total energy of the lattice divided by the number of sites.  In a similar form, we have the magnetic susceptibility

\begin{equation}
\label{e_chi}
\begin{split}
\chi &= \frac{1}{k_B T} \left( \langle M^2 \rangle - \langle M \rangle^2 \right) \\
&= \frac{N_{\text{sites}}^2}{k_B T} \left( \langle m^2 \rangle - \langle m \rangle ^2 \right),
\end{split}
\end{equation}
where $k_B=1$, $T$ is temperature, $M$ is the total magnetization of the lattice, $N_{\text{sites}}$ is the total number of sites, and $m$ is the total magnetization of the lattice divided by the number of sites.  

\begin{figure}
\centering
\includegraphics[width=\linewidth]{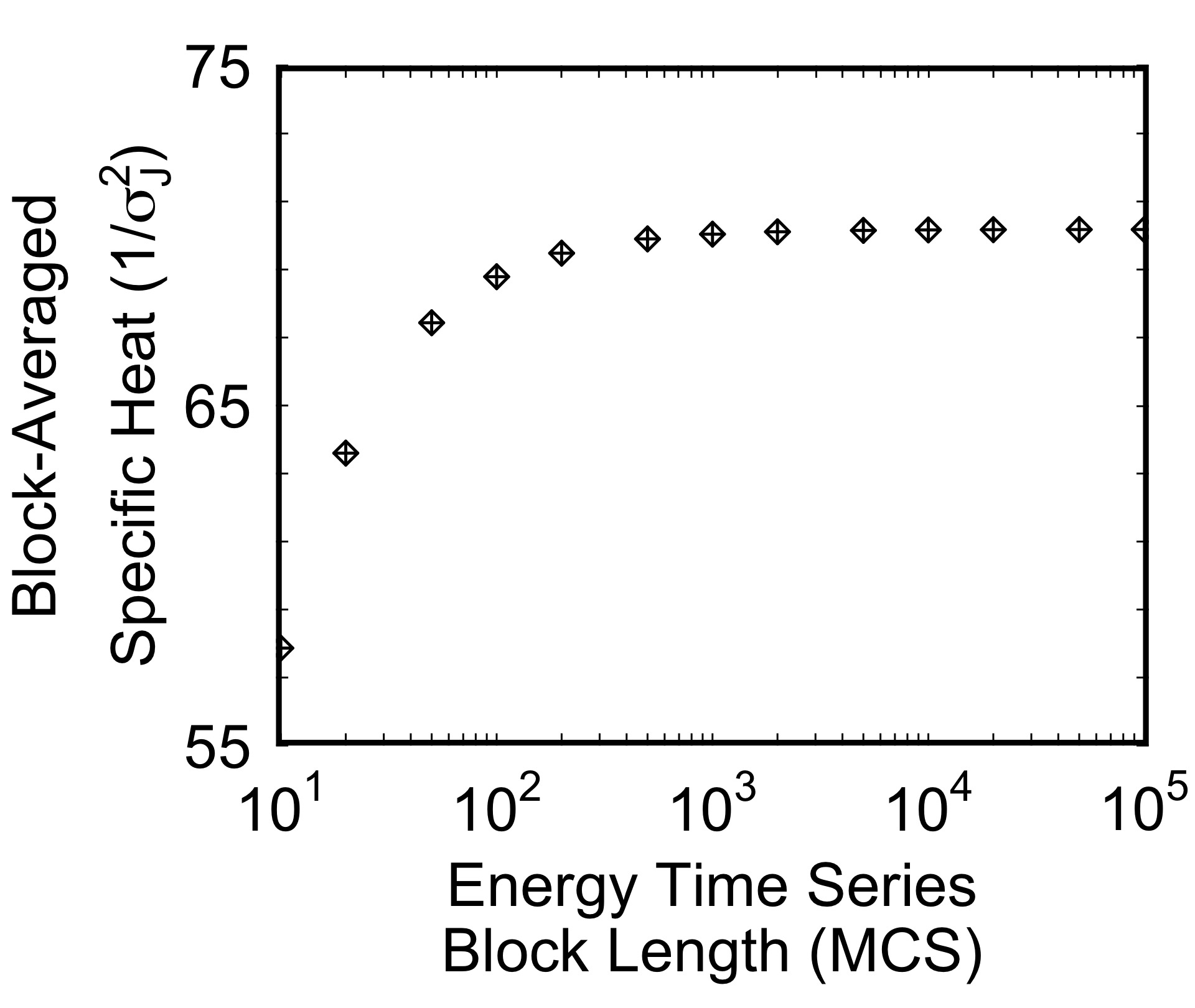}
\caption{Block-averaged specific heat versus energy time series block length for the 2D  ($16 \times 16$) Ising spin glass at $T=2$ averaged over 200 runs.  For longer time series, the specific heat approaches a constant value.}
\label{f-spin-glass-block-specific-heat}
\end{figure}

To calculate the block-averaged specific heat, the energy time series of the system is divided into equally-sized blocks.  The specific heat is calculated for each block using Eq.~(\ref{e_cv}), and then the blocks are averaged together.  For larger block sizes, the specific heat increases and eventually levels off.  The block-averaged specific heat is shown in Fig.~\ref{f-spin-glass-block-specific-heat}.

\begin{figure}
\centering
\includegraphics[width=\linewidth]{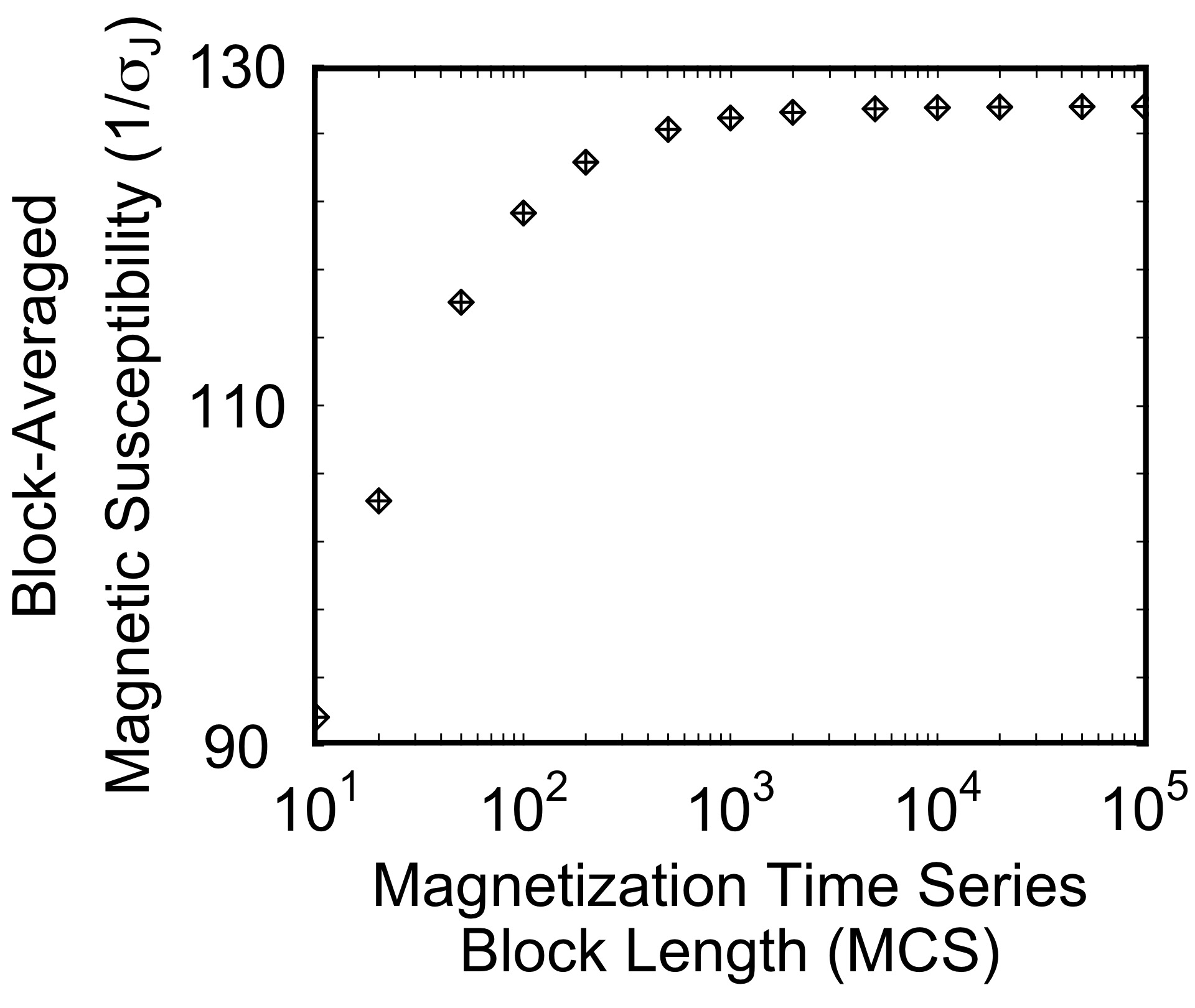}
\caption{Block-averaged magnetic susceptibility versus magnetization time series block length for the 2D  ($16 \times 16$) Ising spin glass at $T=2$ averaged over 200 runs.  For longer time series, the magnetic susceptibility approaches a constant value.}
\label{f-spin-glass-block-susceptibility}
\end{figure}

To calculate the block-averaged magnetic susceptibility, we follow a similar procedure, but we use Eq.~(\ref{e_chi}) with the magnetization time series.  The block-averaged magnetic susceptibility is shown in Fig.~\ref{f-spin-glass-block-susceptibility}.

We can see qualitatively that the results of Figs.~\ref{f-spin-glass-block-specific-heat} and \ref{f-spin-glass-block-susceptibility} are consistent with the spin glass equilibration results of Fig.~\ref{f-spin-glass-susceptibility}.  They indicate that the system is in equilibrium after ${\approx 10^3 \text{ MCS}}$.

\section{Dipoles Aligned Along the X, Y, or Z Axes}
\label{a_xyz-dipoles}

We perform another 200-run set of Monte Carlo simulations of 2D ($16 \times 16$) Ising spins.  In this case, the electric dipoles lie along the x, y, or z axis.  The equilibration times are shown in Table~\ref{t_times2}.

\begin{table}[h!]
	\setlength{\tabcolsep}{6pt}
	\renewcommand{\arraystretch}{1.1}
	\centering
	\begin{tabular}{ |m{2.3cm}|m{3.6cm}| } 
		\hline
		 & Equilibration and \\
		Temperature & Recording Times (MCS)  \\
		\hline
		$1.05 \leq T \leq 10$  & $10^5$ \\
		\hline
		$0.85 \leq T \leq 1$  & $3 \times 10^5$ \\
		\hline
		$0.7 \leq T \leq 0.8$  & $10^6$ \\
		\hline
		$0.6 \leq T \leq 0.65$  & $10^7$ \\
		\hline
	\end{tabular}
	\caption{Equilibration and recording times for the 2D ($16 \times 16$) Ising spin glass with dipoles that lie along the x, y, or z axis for $0.6 \leq T \leq 10$.}
	\label{t_times2}
\end{table}

\begin{figure}
\centering
\includegraphics[width=0.9 \linewidth]{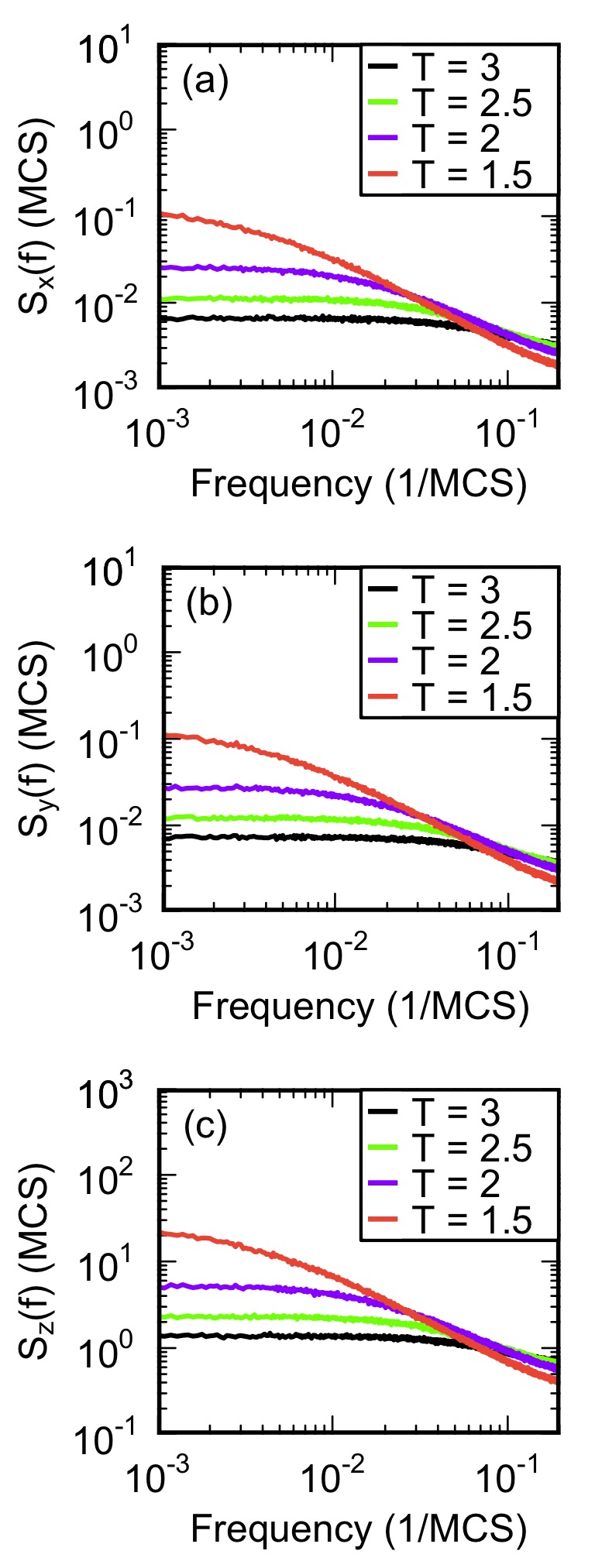}
\caption{Log-log plots of the electric potential noise power versus frequency at QD 1
for fluctuating dipoles that lie along the (a) x-axis, (b) y-axis, and 
(c) z-axis.  All plots are the result of averaging over 200 runs with a QD 
separation of 10 lattice spacings.}
\label{f-panel1}
\end{figure}

The noise power as a function of frequency are shown in Fig.~\ref{f-panel1} for electric dipoles aligned along the (a) x-axis, (b) y-axis, and (c) z-axis.

\begin{figure}
\centering
\includegraphics[width=0.9 \linewidth]{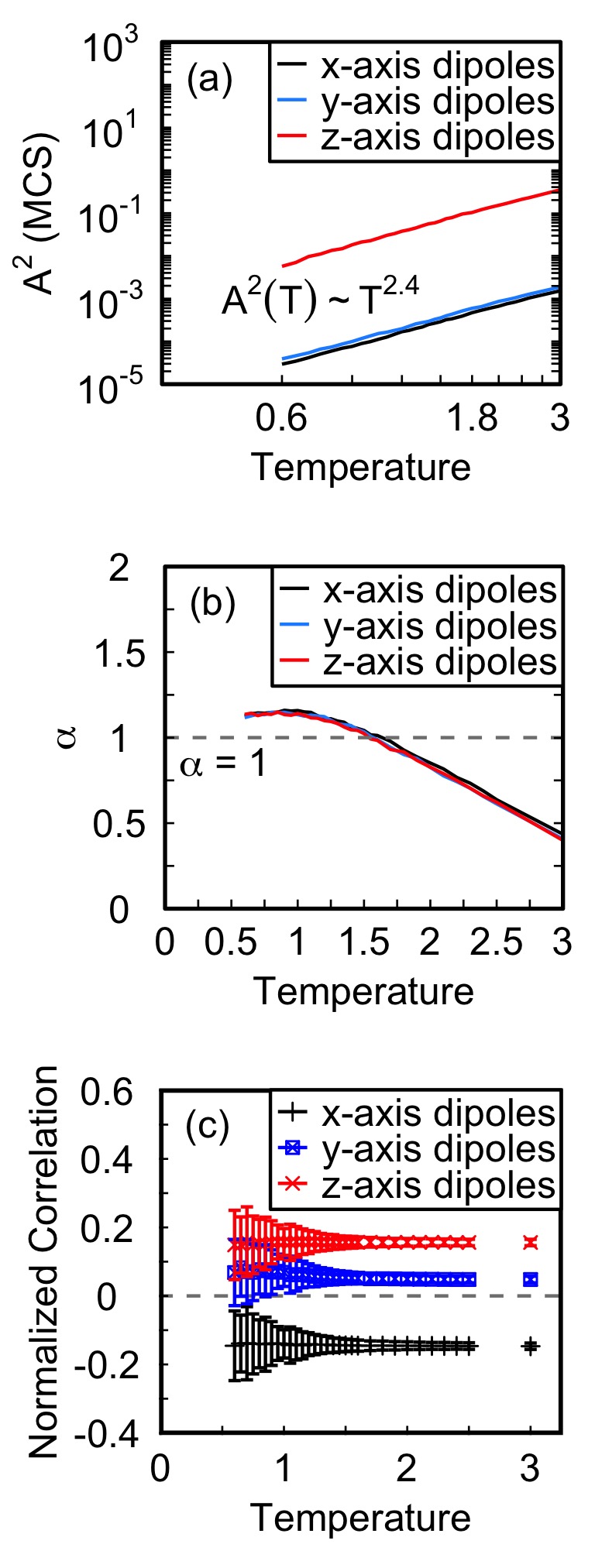}
\caption{(a) Log-log plot of temperature dependence of the noise 
amplitude, $A^2(T)$, obtained from fits to the noise power.  (b) Plot of the temperature dependence of the noise exponent, $\alpha(T)$, obtained from fits to the noise power. (c) Plots of noise correlation between QDs 1 and 2 vs. temperature calculated using Eq. (\ref{e-correlation}). The error bars correspond to the standard deviation that is calculated as described in the main text. Plots (a)-(c) show results for dipoles lying along the x, y, and z axes. All plots are the result of averaging over 200 runs with QD separation of 10 lattice spacings.}
\label{f-panel2}
\end{figure}

$A^2(T)/f^{\alpha(T)}$ is fit to the region of the power spectra that is linear on a log-log plot.  The noise amplitudes ($A^2(T)$) and exponents ($\alpha(T)$) are shown in Fig.~\ref{f-panel2} (a) and (b), respectively.  The dipole correlations are calculated for each axis using Eq.~(\ref{e-correlation}), and they are shown in Fig.~\ref{f-panel2} (c).

The frequency-dependent correlation and phase for the fixed-axis dipoles are calculated using Eq.~(\ref{e-correlationf}).  The correlation and phase are shown in Fig.~\ref{f-correlationf}.  For y-axis dipoles, the factor $\vec{p}_{\eta,i}(t) \cdot \hat{R}_\eta$ from Eq.~(\ref{eq:Vpotential}) is small for fluctuators close to the line connecting the two quantum dots.  $\vec{p}_{\eta,i}(t) \cdot \hat{R}_\eta=0$ for y-axis dipoles along this line.  From this, we would expect the correlation strength to be smallest for y-axis dipoles, and this is seen in Fig.~\ref{f-correlationf}(a).  As seen in Fig.~\ref{f-correlationf}(b), the dipole potentials at quantum dots 1 and 2 for x-axis dipoles are $\pi$ radians out of phase.  Since the quantum dots lie along the x-axis, any fluctuations in x-axis dipoles between the quantum dots will have the opposite effect on each dot.

The image dipoles are aligned for the component along the z-axis and antialigned for the components along the x and y axes (see Fig. \ref{f-Aa2}), so we expect the dipole potential calculated from the z-axis component to be the most significant.  For the electric dipoles with random orientations, the results should be similar to the case of dipoles aligned along the z-axis.  This is certainly true for the dipole correlations and the noise exponents.  However, the noise power and noise amplitudes resulting from electric dipoles aligned along the z-axis are three times larger.  This is expected, since the z-axis dipoles have unit length, but the z components of the randomly-oriented dipoles have a typical length of $1/\sqrt{3}$.  This corresponds to a factor of $1/\sqrt{3}^2=1/3$ that reduces the noise amplitude.

\clearpage


\begin{thebibliography}{59}%
\makeatletter
\providecommand \@ifxundefined [1]{%
 \@ifx{#1\undefined}
}%
\providecommand \@ifnum [1]{%
 \ifnum #1\expandafter \@firstoftwo
 \else \expandafter \@secondoftwo
 \fi
}%
\providecommand \@ifx [1]{%
 \ifx #1\expandafter \@firstoftwo
 \else \expandafter \@secondoftwo
 \fi
}%
\providecommand \natexlab [1]{#1}%
\providecommand \enquote  [1]{``#1''}%
\providecommand \bibnamefont  [1]{#1}%
\providecommand \bibfnamefont [1]{#1}%
\providecommand \citenamefont [1]{#1}%
\providecommand \href@noop [0]{\@secondoftwo}%
\providecommand \href [0]{\begingroup \@sanitize@url \@href}%
\providecommand \@href[1]{\@@startlink{#1}\@@href}%
\providecommand \@@href[1]{\endgroup#1\@@endlink}%
\providecommand \@sanitize@url [0]{\catcode `\\12\catcode `\$12\catcode
  `\&12\catcode `\#12\catcode `\^12\catcode `\_12\catcode `\%12\relax}%
\providecommand \@@startlink[1]{}%
\providecommand \@@endlink[0]{}%
\providecommand \url  [0]{\begingroup\@sanitize@url \@url }%
\providecommand \@url [1]{\endgroup\@href {#1}{\urlprefix }}%
\providecommand \urlprefix  [0]{URL }%
\providecommand \Eprint [0]{\href }%
\providecommand \doibase [0]{https://doi.org/}%
\providecommand \selectlanguage [0]{\@gobble}%
\providecommand \bibinfo  [0]{\@secondoftwo}%
\providecommand \bibfield  [0]{\@secondoftwo}%
\providecommand \translation [1]{[#1]}%
\providecommand \BibitemOpen [0]{}%
\providecommand \bibitemStop [0]{}%
\providecommand \bibitemNoStop [0]{.\EOS\space}%
\providecommand \EOS [0]{\spacefactor3000\relax}%
\providecommand \BibitemShut  [1]{\csname bibitem#1\endcsname}%
\let\auto@bib@innerbib\@empty
\bibitem [{\citenamefont {Yoneda}\ \emph {et~al.}(2018)\citenamefont {Yoneda},
  \citenamefont {Takeda}, \citenamefont {Otsuka}, \citenamefont {Nakajima},
  \citenamefont {Delbecq}, \citenamefont {Allison}, \citenamefont {Honda},
  \citenamefont {Kodera}, \citenamefont {Oda}, \citenamefont {Hoshi},
  \citenamefont {Usami}, \citenamefont {Itoh},\ and\ \citenamefont
  {Tarucha}}]{Yoneda2018}%
  \BibitemOpen
  \bibfield  {author} {\bibinfo {author} {\bibfnamefont {J.}~\bibnamefont
  {Yoneda}}, \bibinfo {author} {\bibfnamefont {K.}~\bibnamefont {Takeda}},
  \bibinfo {author} {\bibfnamefont {T.}~\bibnamefont {Otsuka}}, \bibinfo
  {author} {\bibfnamefont {T.}~\bibnamefont {Nakajima}}, \bibinfo {author}
  {\bibfnamefont {M.~R.}\ \bibnamefont {Delbecq}}, \bibinfo {author}
  {\bibfnamefont {G.}~\bibnamefont {Allison}}, \bibinfo {author} {\bibfnamefont
  {T.}~\bibnamefont {Honda}}, \bibinfo {author} {\bibfnamefont
  {T.}~\bibnamefont {Kodera}}, \bibinfo {author} {\bibfnamefont
  {S.}~\bibnamefont {Oda}}, \bibinfo {author} {\bibfnamefont {Y.}~\bibnamefont
  {Hoshi}}, \bibinfo {author} {\bibfnamefont {N.}~\bibnamefont {Usami}},
  \bibinfo {author} {\bibfnamefont {K.~M.}\ \bibnamefont {Itoh}},\ and\
  \bibinfo {author} {\bibfnamefont {S.}~\bibnamefont {Tarucha}},\ }\bibfield
  {title} {\bibinfo {title} {A quantum-dot spin qubit with coherence limited by
  charge noise and fidelity higher than 99.9\%},\ }\href
  {https://doi.org/10.1038/s41565-017-0014-x} {\bibfield  {journal} {\bibinfo
  {journal} {Nature Nanotechnology}\ }\textbf {\bibinfo {volume} {13}},\
  \bibinfo {pages} {102} (\bibinfo {year} {2018})}\BibitemShut {NoStop}%
\bibitem [{\citenamefont {Connors}\ \emph {et~al.}(2022)\citenamefont
  {Connors}, \citenamefont {Nelson},\ and\ \citenamefont
  {Nichol}}]{Connors2022}%
  \BibitemOpen
  \bibfield  {author} {\bibinfo {author} {\bibfnamefont {E.~J.}\ \bibnamefont
  {Connors}}, \bibinfo {author} {\bibfnamefont {J.}~\bibnamefont {Nelson}},\
  and\ \bibinfo {author} {\bibfnamefont {J.~M.}\ \bibnamefont {Nichol}},\
  }\bibfield  {title} {\bibinfo {title} {Charge-noise spectroscopy of {Si/SiGe}
  quantum dots via dynamically-decoupled exchange oscillations},\ }\href
  {https://doi.org/10.1038/s41467-022-28519-x} {\bibfield  {journal} {\bibinfo
  {journal} {Nature Communications}\ }\textbf {\bibinfo {volume} {13}},\
  \bibinfo {pages} {940} (\bibinfo {year} {2022})}\BibitemShut {NoStop}%
\bibitem [{\citenamefont {Dutta}\ \emph {et~al.}(1979)\citenamefont {Dutta},
  \citenamefont {Dimon},\ and\ \citenamefont {Horn}}]{Dutta1979}%
  \BibitemOpen
  \bibfield  {author} {\bibinfo {author} {\bibfnamefont {P.}~\bibnamefont
  {Dutta}}, \bibinfo {author} {\bibfnamefont {P.}~\bibnamefont {Dimon}},\ and\
  \bibinfo {author} {\bibfnamefont {P.~M.}\ \bibnamefont {Horn}},\ }\bibfield
  {title} {\bibinfo {title} {Energy scales for noise processes in metals},\
  }\href {https://doi.org/10.1103/PhysRevLett.43.646} {\bibfield  {journal}
  {\bibinfo  {journal} {Phys. Rev. Lett.}\ }\textbf {\bibinfo {volume} {43}},\
  \bibinfo {pages} {646} (\bibinfo {year} {1979})}\BibitemShut {NoStop}%
\bibitem [{\citenamefont {Dutta}\ and\ \citenamefont {Horn}(1981)}]{Dutta1981}%
  \BibitemOpen
  \bibfield  {author} {\bibinfo {author} {\bibfnamefont {P.}~\bibnamefont
  {Dutta}}\ and\ \bibinfo {author} {\bibfnamefont {P.~M.}\ \bibnamefont
  {Horn}},\ }\bibfield  {title} {\bibinfo {title} {Low-frequency fluctuations
  in solids: $\frac{1}{f}$ noise},\ }\href
  {https://doi.org/10.1103/RevModPhys.53.497} {\bibfield  {journal} {\bibinfo
  {journal} {Rev. Mod. Phys.}\ }\textbf {\bibinfo {volume} {53}},\ \bibinfo
  {pages} {497} (\bibinfo {year} {1981})}\BibitemShut {NoStop}%
\bibitem [{\citenamefont {Petit}\ \emph {et~al.}(2020)\citenamefont {Petit},
  \citenamefont {Eenink}, \citenamefont {Russ}, \citenamefont {Lawrie},
  \citenamefont {Hendrickx}, \citenamefont {Philips}, \citenamefont {Clarke},
  \citenamefont {Vandersypen},\ and\ \citenamefont {Veldhorst}}]{Petit2020}%
  \BibitemOpen
  \bibfield  {author} {\bibinfo {author} {\bibfnamefont {L.}~\bibnamefont
  {Petit}}, \bibinfo {author} {\bibfnamefont {H.~G.~J.}\ \bibnamefont
  {Eenink}}, \bibinfo {author} {\bibfnamefont {M.}~\bibnamefont {Russ}},
  \bibinfo {author} {\bibfnamefont {W.~I.~L.}\ \bibnamefont {Lawrie}}, \bibinfo
  {author} {\bibfnamefont {N.~W.}\ \bibnamefont {Hendrickx}}, \bibinfo {author}
  {\bibfnamefont {S.~G.~J.}\ \bibnamefont {Philips}}, \bibinfo {author}
  {\bibfnamefont {J.~S.}\ \bibnamefont {Clarke}}, \bibinfo {author}
  {\bibfnamefont {L.~M.~K.}\ \bibnamefont {Vandersypen}},\ and\ \bibinfo
  {author} {\bibfnamefont {M.}~\bibnamefont {Veldhorst}},\ }\bibfield  {title}
  {\bibinfo {title} {Universal quantum logic in hot silicon qubits},\
  }\href@noop {} {\bibfield  {journal} {\bibinfo  {journal} {Nature}\ }\textbf
  {\bibinfo {volume} {580}},\ \bibinfo {pages} {355} (\bibinfo {year}
  {2020})}\BibitemShut {NoStop}%
\bibitem [{\citenamefont {Connors}\ \emph {et~al.}(2019)\citenamefont
  {Connors}, \citenamefont {Nelson}, \citenamefont {Qiao}, \citenamefont
  {Edge},\ and\ \citenamefont {Nichol}}]{Connors2019}%
  \BibitemOpen
  \bibfield  {author} {\bibinfo {author} {\bibfnamefont {E.~J.}\ \bibnamefont
  {Connors}}, \bibinfo {author} {\bibfnamefont {J.}~\bibnamefont {Nelson}},
  \bibinfo {author} {\bibfnamefont {H.}~\bibnamefont {Qiao}}, \bibinfo {author}
  {\bibfnamefont {L.~F.}\ \bibnamefont {Edge}},\ and\ \bibinfo {author}
  {\bibfnamefont {J.~M.}\ \bibnamefont {Nichol}},\ }\bibfield  {title}
  {\bibinfo {title} {Low-frequency charge noise in {Si/SiGe} quantum dots},\
  }\href {https://doi.org/10.1103/PhysRevB.100.165305} {\bibfield  {journal}
  {\bibinfo  {journal} {Phys. Rev. B}\ }\textbf {\bibinfo {volume} {100}},\
  \bibinfo {pages} {165305} (\bibinfo {year} {2019})}\BibitemShut {NoStop}%
\bibitem [{\citenamefont {Connors}\ \emph {et~al.}(2020)\citenamefont
  {Connors}, \citenamefont {Nelson}, \citenamefont {Qiao}, \citenamefont
  {Edge},\ and\ \citenamefont {Nichol}}]{Connors2019e}%
  \BibitemOpen
  \bibfield  {author} {\bibinfo {author} {\bibfnamefont {E.~J.}\ \bibnamefont
  {Connors}}, \bibinfo {author} {\bibfnamefont {J.}~\bibnamefont {Nelson}},
  \bibinfo {author} {\bibfnamefont {H.}~\bibnamefont {Qiao}}, \bibinfo {author}
  {\bibfnamefont {L.~F.}\ \bibnamefont {Edge}},\ and\ \bibinfo {author}
  {\bibfnamefont {J.~M.}\ \bibnamefont {Nichol}},\ }\bibfield  {title}
  {\bibinfo {title} {Erratum: Low-frequency charge noise in {Si/SiGe} quantum
  dots {[Phys. Rev. B 100, 165305 (2019)]}},\ }\href
  {https://doi.org/10.1103/PhysRevB.102.039902} {\bibfield  {journal} {\bibinfo
   {journal} {Phys. Rev. B}\ }\textbf {\bibinfo {volume} {102}},\ \bibinfo
  {pages} {039902(E)} (\bibinfo {year} {2020})}\BibitemShut {NoStop}%
\bibitem [{\citenamefont {Petit}\ \emph {et~al.}(2018)\citenamefont {Petit},
  \citenamefont {Boter}, \citenamefont {Eenink}, \citenamefont {Droulers},
  \citenamefont {Tagliaferri}, \citenamefont {Li}, \citenamefont {Franke},
  \citenamefont {Singh}, \citenamefont {Clarke}, \citenamefont {Schouten},
  \citenamefont {Dobrovitski}, \citenamefont {Vandersypen},\ and\ \citenamefont
  {Veldhorst}}]{Petit2018}%
  \BibitemOpen
  \bibfield  {author} {\bibinfo {author} {\bibfnamefont {L.}~\bibnamefont
  {Petit}}, \bibinfo {author} {\bibfnamefont {J.~M.}\ \bibnamefont {Boter}},
  \bibinfo {author} {\bibfnamefont {H.~G.~J.}\ \bibnamefont {Eenink}}, \bibinfo
  {author} {\bibfnamefont {G.}~\bibnamefont {Droulers}}, \bibinfo {author}
  {\bibfnamefont {M.~L.~V.}\ \bibnamefont {Tagliaferri}}, \bibinfo {author}
  {\bibfnamefont {R.}~\bibnamefont {Li}}, \bibinfo {author} {\bibfnamefont
  {D.~P.}\ \bibnamefont {Franke}}, \bibinfo {author} {\bibfnamefont {K.~J.}\
  \bibnamefont {Singh}}, \bibinfo {author} {\bibfnamefont {J.~S.}\ \bibnamefont
  {Clarke}}, \bibinfo {author} {\bibfnamefont {R.~N.}\ \bibnamefont
  {Schouten}}, \bibinfo {author} {\bibfnamefont {V.~V.}\ \bibnamefont
  {Dobrovitski}}, \bibinfo {author} {\bibfnamefont {L.~M.~K.}\ \bibnamefont
  {Vandersypen}},\ and\ \bibinfo {author} {\bibfnamefont {M.}~\bibnamefont
  {Veldhorst}},\ }\bibfield  {title} {\bibinfo {title} {Spin lifetime and
  charge noise in hot silicon quantum dot qubits},\ }\href
  {https://doi.org/10.1103/PhysRevLett.121.076801} {\bibfield  {journal}
  {\bibinfo  {journal} {Phys. Rev. Lett.}\ }\textbf {\bibinfo {volume} {121}},\
  \bibinfo {pages} {076801} (\bibinfo {year} {2018})}\BibitemShut {NoStop}%
\bibitem [{\citenamefont {Spence}\ \emph {et~al.}(2023)\citenamefont {Spence},
  \citenamefont {Cardoso~Paz}, \citenamefont {Michal}, \citenamefont
  {Chanrion}, \citenamefont {Niegemann}, \citenamefont {Jadot}, \citenamefont
  {Mortemousque}, \citenamefont {Klemt}, \citenamefont {Thiney}, \citenamefont
  {Bertrand}, \citenamefont {Hutin}, \citenamefont {B\"auerle}, \citenamefont
  {Vinet}, \citenamefont {Niquet}, \citenamefont {Meunier},\ and\ \citenamefont
  {Urdampilleta}}]{Spence2023}%
  \BibitemOpen
  \bibfield  {author} {\bibinfo {author} {\bibfnamefont {C.}~\bibnamefont
  {Spence}}, \bibinfo {author} {\bibfnamefont {B.}~\bibnamefont {Cardoso~Paz}},
  \bibinfo {author} {\bibfnamefont {V.}~\bibnamefont {Michal}}, \bibinfo
  {author} {\bibfnamefont {E.}~\bibnamefont {Chanrion}}, \bibinfo {author}
  {\bibfnamefont {D.~J.}\ \bibnamefont {Niegemann}}, \bibinfo {author}
  {\bibfnamefont {B.}~\bibnamefont {Jadot}}, \bibinfo {author} {\bibfnamefont
  {P.-A.}\ \bibnamefont {Mortemousque}}, \bibinfo {author} {\bibfnamefont
  {B.}~\bibnamefont {Klemt}}, \bibinfo {author} {\bibfnamefont
  {V.}~\bibnamefont {Thiney}}, \bibinfo {author} {\bibfnamefont
  {B.}~\bibnamefont {Bertrand}}, \bibinfo {author} {\bibfnamefont
  {L.}~\bibnamefont {Hutin}}, \bibinfo {author} {\bibfnamefont
  {C.}~\bibnamefont {B\"auerle}}, \bibinfo {author} {\bibfnamefont
  {M.}~\bibnamefont {Vinet}}, \bibinfo {author} {\bibfnamefont {Y.-M.}\
  \bibnamefont {Niquet}}, \bibinfo {author} {\bibfnamefont {T.}~\bibnamefont
  {Meunier}},\ and\ \bibinfo {author} {\bibfnamefont {M.}~\bibnamefont
  {Urdampilleta}},\ }\bibfield  {title} {\bibinfo {title} {Probing
  low-frequency charge noise in few-electron {CMOS} quantum dots},\ }\href
  {https://doi.org/10.1103/PhysRevApplied.19.044010} {\bibfield  {journal}
  {\bibinfo  {journal} {Phys. Rev. Appl.}\ }\textbf {\bibinfo {volume} {19}},\
  \bibinfo {pages} {044010} (\bibinfo {year} {2023})}\BibitemShut {NoStop}%
\bibitem [{\citenamefont {Ahn}\ \emph {et~al.}(2021)\citenamefont {Ahn},
  \citenamefont {Das~Sarma},\ and\ \citenamefont {Kestner}}]{Ahn2021}%
  \BibitemOpen
  \bibfield  {author} {\bibinfo {author} {\bibfnamefont {S.}~\bibnamefont
  {Ahn}}, \bibinfo {author} {\bibfnamefont {S.}~\bibnamefont {Das~Sarma}},\
  and\ \bibinfo {author} {\bibfnamefont {J.~P.}\ \bibnamefont {Kestner}},\
  }\bibfield  {title} {\bibinfo {title} {Microscopic bath effects on noise
  spectra in semiconductor quantum dot qubits},\ }\href
  {https://doi.org/10.1103/PhysRevB.103.L041304} {\bibfield  {journal}
  {\bibinfo  {journal} {Phys. Rev. B}\ }\textbf {\bibinfo {volume} {103}},\
  \bibinfo {pages} {L041304} (\bibinfo {year} {2021})}\BibitemShut {NoStop}%
\bibitem [{\citenamefont {Throckmorton}\ and\ \citenamefont
  {Sarma}(2023)}]{Throckmorton2023}%
  \BibitemOpen
  \bibfield  {author} {\bibinfo {author} {\bibfnamefont {R.~E.}\ \bibnamefont
  {Throckmorton}}\ and\ \bibinfo {author} {\bibfnamefont {S.~D.}\ \bibnamefont
  {Sarma}},\ }\href@noop {} {\bibinfo {title} {A generalized model of the noise
  spectrum of a two-level fluctuator in the presence of an electron subbath}}
  (\bibinfo {year} {2023}),\ \Eprint {https://arxiv.org/abs/2305.14348}
  {arXiv:2305.14348 [cond-mat.mes-hall]} \BibitemShut {NoStop}%
\bibitem [{\citenamefont {Mickelsen}\ \emph {et~al.}(2023)\citenamefont
  {Mickelsen}, \citenamefont {Carruzzo}, \citenamefont {Coppersmith},\ and\
  \citenamefont {Yu}}]{Mickelsen2023}%
  \BibitemOpen
  \bibfield  {author} {\bibinfo {author} {\bibfnamefont {D.~L.}\ \bibnamefont
  {Mickelsen}}, \bibinfo {author} {\bibfnamefont {H.~M.}\ \bibnamefont
  {Carruzzo}}, \bibinfo {author} {\bibfnamefont {S.~N.}\ \bibnamefont
  {Coppersmith}},\ and\ \bibinfo {author} {\bibfnamefont {C.~C.}\ \bibnamefont
  {Yu}},\ }\bibfield  {title} {\bibinfo {title} {Effects of temperature
  fluctuations on charge noise in quantum dot qubits},\ }\href
  {https://doi.org/10.1103/PhysRevB.108.075303} {\bibfield  {journal} {\bibinfo
   {journal} {Phys. Rev. B}\ }\textbf {\bibinfo {volume} {108}},\ \bibinfo
  {pages} {075303} (\bibinfo {year} {2023})}\BibitemShut {NoStop}%
\bibitem [{\citenamefont {Yoneda}\ \emph {et~al.}(2023)\citenamefont {Yoneda},
  \citenamefont {Rojas-Arias}, \citenamefont {Stano}, \citenamefont {Takeda},
  \citenamefont {Noiri}, \citenamefont {Nakajima}, \citenamefont {Loss},\ and\
  \citenamefont {Tarucha}}]{Yoneda2023}%
  \BibitemOpen
  \bibfield  {author} {\bibinfo {author} {\bibfnamefont {J.}~\bibnamefont
  {Yoneda}}, \bibinfo {author} {\bibfnamefont {J.~S.}\ \bibnamefont
  {Rojas-Arias}}, \bibinfo {author} {\bibfnamefont {P.}~\bibnamefont {Stano}},
  \bibinfo {author} {\bibfnamefont {K.}~\bibnamefont {Takeda}}, \bibinfo
  {author} {\bibfnamefont {A.}~\bibnamefont {Noiri}}, \bibinfo {author}
  {\bibfnamefont {T.}~\bibnamefont {Nakajima}}, \bibinfo {author}
  {\bibfnamefont {D.}~\bibnamefont {Loss}},\ and\ \bibinfo {author}
  {\bibfnamefont {S.}~\bibnamefont {Tarucha}},\ }\bibfield  {title} {\bibinfo
  {title} {Noise-correlation spectrum for a pair of spin qubits in silicon},\
  }\bibfield  {journal} {\bibinfo  {journal} {Nature Phys.}\ }\href
  {https://doi.org/10.1038/s41567-023-02238-6} {10.1038/s41567-023-02238-6}
  (\bibinfo {year} {2023})\BibitemShut {NoStop}%
\bibitem [{\citenamefont {Rojas-Arias}\ \emph {et~al.}(2023)\citenamefont
  {Rojas-Arias}, \citenamefont {Noiri}, \citenamefont {Stano}, \citenamefont
  {Nakajima}, \citenamefont {Yoneda}, \citenamefont {Takeda}, \citenamefont
  {Kobayashi}, \citenamefont {Sammak}, \citenamefont {Scappucci}, \citenamefont
  {Loss},\ and\ \citenamefont {Tarucha}}]{RojasArias2023}%
  \BibitemOpen
  \bibfield  {author} {\bibinfo {author} {\bibfnamefont {J.~S.}\ \bibnamefont
  {Rojas-Arias}}, \bibinfo {author} {\bibfnamefont {A.}~\bibnamefont {Noiri}},
  \bibinfo {author} {\bibfnamefont {P.}~\bibnamefont {Stano}}, \bibinfo
  {author} {\bibfnamefont {T.}~\bibnamefont {Nakajima}}, \bibinfo {author}
  {\bibfnamefont {J.}~\bibnamefont {Yoneda}}, \bibinfo {author} {\bibfnamefont
  {K.}~\bibnamefont {Takeda}}, \bibinfo {author} {\bibfnamefont
  {T.}~\bibnamefont {Kobayashi}}, \bibinfo {author} {\bibfnamefont
  {A.}~\bibnamefont {Sammak}}, \bibinfo {author} {\bibfnamefont
  {G.}~\bibnamefont {Scappucci}}, \bibinfo {author} {\bibfnamefont
  {D.}~\bibnamefont {Loss}},\ and\ \bibinfo {author} {\bibfnamefont
  {S.}~\bibnamefont {Tarucha}},\ }\href@noop {} {\bibinfo {title} {Spatial
  noise correlations beyond nearest-neighbor in {${}^{28}$Si/SiGe} spin
  qubits}} (\bibinfo {year} {2023}),\ \Eprint
  {https://arxiv.org/abs/2302.11717} {arXiv:2302.11717 [cond-mat.mes-hall]}
  \BibitemShut {NoStop}%
\bibitem [{\citenamefont {Hu}\ and\ \citenamefont {Das~Sarma}(2006)}]{Hu2006}%
  \BibitemOpen
  \bibfield  {author} {\bibinfo {author} {\bibfnamefont {X.}~\bibnamefont
  {Hu}}\ and\ \bibinfo {author} {\bibfnamefont {S.}~\bibnamefont {Das~Sarma}},\
  }\bibfield  {title} {\bibinfo {title} {Charge-fluctuation-induced dephasing
  of exchange-coupled spin qubits},\ }\href
  {https://doi.org/10.1103/PhysRevLett.96.100501} {\bibfield  {journal}
  {\bibinfo  {journal} {Phys. Rev. Lett.}\ }\textbf {\bibinfo {volume} {96}},\
  \bibinfo {pages} {100501} (\bibinfo {year} {2006})}\BibitemShut {NoStop}%
\bibitem [{\citenamefont {Culcer}\ \emph {et~al.}(2009)\citenamefont {Culcer},
  \citenamefont {Hu},\ and\ \citenamefont {Das~Sarma}}]{Culcer2009}%
  \BibitemOpen
  \bibfield  {author} {\bibinfo {author} {\bibfnamefont {D.}~\bibnamefont
  {Culcer}}, \bibinfo {author} {\bibfnamefont {X.}~\bibnamefont {Hu}},\ and\
  \bibinfo {author} {\bibfnamefont {S.}~\bibnamefont {Das~Sarma}},\ }\bibfield
  {title} {\bibinfo {title} {{Dephasing of {Si} spin qubits due to charge
  noise}},\ }\href {https://doi.org/10.1063/1.3194778} {\bibfield  {journal}
  {\bibinfo  {journal} {Applied Physics Letters}\ }\textbf {\bibinfo {volume}
  {95}},\ \bibinfo {pages} {073102} (\bibinfo {year} {2009})}\BibitemShut
  {NoStop}%
\bibitem [{\citenamefont {Ramon}\ and\ \citenamefont {Hu}(2010)}]{Ramon2010}%
  \BibitemOpen
  \bibfield  {author} {\bibinfo {author} {\bibfnamefont {G.}~\bibnamefont
  {Ramon}}\ and\ \bibinfo {author} {\bibfnamefont {X.}~\bibnamefont {Hu}},\
  }\bibfield  {title} {\bibinfo {title} {Decoherence of spin qubits due to a
  nearby charge fluctuator in gate-defined double dots},\ }\href
  {https://doi.org/10.1103/PhysRevB.81.045304} {\bibfield  {journal} {\bibinfo
  {journal} {Phys. Rev. B}\ }\textbf {\bibinfo {volume} {81}},\ \bibinfo
  {pages} {045304} (\bibinfo {year} {2010})}\BibitemShut {NoStop}%
\bibitem [{\citenamefont {Li}\ \emph {et~al.}(2010)\citenamefont {Li},
  \citenamefont {Cywi\ifmmode~\acute{n}\else \'{n}\fi{}ski}, \citenamefont
  {Culcer}, \citenamefont {Hu},\ and\ \citenamefont {Das~Sarma}}]{Li2010}%
  \BibitemOpen
  \bibfield  {author} {\bibinfo {author} {\bibfnamefont {Q.}~\bibnamefont
  {Li}}, \bibinfo {author} {\bibfnamefont {L.}~\bibnamefont
  {Cywi\ifmmode~\acute{n}\else \'{n}\fi{}ski}}, \bibinfo {author}
  {\bibfnamefont {D.}~\bibnamefont {Culcer}}, \bibinfo {author} {\bibfnamefont
  {X.}~\bibnamefont {Hu}},\ and\ \bibinfo {author} {\bibfnamefont
  {S.}~\bibnamefont {Das~Sarma}},\ }\bibfield  {title} {\bibinfo {title}
  {Exchange coupling in silicon quantum dots: Theoretical considerations for
  quantum computation},\ }\href {https://doi.org/10.1103/PhysRevB.81.085313}
  {\bibfield  {journal} {\bibinfo  {journal} {Phys. Rev. B}\ }\textbf {\bibinfo
  {volume} {81}},\ \bibinfo {pages} {085313} (\bibinfo {year}
  {2010})}\BibitemShut {NoStop}%
\bibitem [{\citenamefont {Culcer}\ and\ \citenamefont
  {Zimmerman}(2013)}]{Culcer2013}%
  \BibitemOpen
  \bibfield  {author} {\bibinfo {author} {\bibfnamefont {D.}~\bibnamefont
  {Culcer}}\ and\ \bibinfo {author} {\bibfnamefont {N.~M.}\ \bibnamefont
  {Zimmerman}},\ }\bibfield  {title} {\bibinfo {title} {{Dephasing of {Si}
  singlet-triplet qubits due to charge and spin defects}},\ }\href
  {https://doi.org/10.1063/1.4810911} {\bibfield  {journal} {\bibinfo
  {journal} {Applied Physics Letters}\ }\textbf {\bibinfo {volume} {102}},\
  \bibinfo {pages} {232108} (\bibinfo {year} {2013})}\BibitemShut {NoStop}%
\bibitem [{\citenamefont {Huang}\ \emph {et~al.}(2018)\citenamefont {Huang},
  \citenamefont {Zimmerman},\ and\ \citenamefont {Bryant}}]{Huang2018}%
  \BibitemOpen
  \bibfield  {author} {\bibinfo {author} {\bibfnamefont {P.}~\bibnamefont
  {Huang}}, \bibinfo {author} {\bibfnamefont {N.~M.}\ \bibnamefont
  {Zimmerman}},\ and\ \bibinfo {author} {\bibfnamefont {G.~W.}\ \bibnamefont
  {Bryant}},\ }\bibfield  {title} {\bibinfo {title} {Spin decoherence in a
  two-qubit cphase gate: the critical role of tunneling noise},\ }\href
  {https://doi.org/10.1038/s41534-018-0112-0} {\bibfield  {journal} {\bibinfo
  {journal} {npj Quantum Inf.}\ }\textbf {\bibinfo {volume} {4}},\ \bibinfo
  {pages} {62} (\bibinfo {year} {2018})}\BibitemShut {NoStop}%
\bibitem [{\citenamefont {Yang}\ and\ \citenamefont {Wang}(2017)}]{Yang2017}%
  \BibitemOpen
  \bibfield  {author} {\bibinfo {author} {\bibfnamefont {X.-C.}\ \bibnamefont
  {Yang}}\ and\ \bibinfo {author} {\bibfnamefont {X.}~\bibnamefont {Wang}},\
  }\bibfield  {title} {\bibinfo {title} {Suppression of charge noise using
  barrier control of a singlet-triplet qubit},\ }\href
  {https://doi.org/10.1103/PhysRevA.96.012318} {\bibfield  {journal} {\bibinfo
  {journal} {Phys. Rev. A}\ }\textbf {\bibinfo {volume} {96}},\ \bibinfo
  {pages} {012318} (\bibinfo {year} {2017})}\BibitemShut {NoStop}%
\bibitem [{\citenamefont {Shim}\ and\ \citenamefont {Tahan}(2018)}]{Shim2018}%
  \BibitemOpen
  \bibfield  {author} {\bibinfo {author} {\bibfnamefont {Y.-P.}\ \bibnamefont
  {Shim}}\ and\ \bibinfo {author} {\bibfnamefont {C.}~\bibnamefont {Tahan}},\
  }\bibfield  {title} {\bibinfo {title} {Barrier versus tilt exchange gate
  operations in spin-based quantum computing},\ }\href
  {https://doi.org/10.1103/PhysRevB.97.155402} {\bibfield  {journal} {\bibinfo
  {journal} {Phys. Rev. B}\ }\textbf {\bibinfo {volume} {97}},\ \bibinfo
  {pages} {155402} (\bibinfo {year} {2018})}\BibitemShut {NoStop}%
\bibitem [{\citenamefont {G\"ung\"ord\"u}\ and\ \citenamefont
  {Kestner}(2019)}]{Gungordu2019}%
  \BibitemOpen
  \bibfield  {author} {\bibinfo {author} {\bibfnamefont {U.}~\bibnamefont
  {G\"ung\"ord\"u}}\ and\ \bibinfo {author} {\bibfnamefont {J.~P.}\
  \bibnamefont {Kestner}},\ }\bibfield  {title} {\bibinfo {title} {Indications
  of a soft cutoff frequency in the charge noise of a {Si/SiGe} quantum dot
  spin qubit},\ }\href {https://doi.org/10.1103/PhysRevB.99.081301} {\bibfield
  {journal} {\bibinfo  {journal} {Phys. Rev. B}\ }\textbf {\bibinfo {volume}
  {99}},\ \bibinfo {pages} {081301} (\bibinfo {year} {2019})}\BibitemShut
  {NoStop}%
\bibitem [{\citenamefont {Shehata}\ \emph {et~al.}(2023)\citenamefont
  {Shehata}, \citenamefont {Simion}, \citenamefont {Li}, \citenamefont
  {Mohiyaddin}, \citenamefont {Wan}, \citenamefont {Mongillo}, \citenamefont
  {Govoreanu}, \citenamefont {Radu}, \citenamefont {De~Greve},\ and\
  \citenamefont {Van~Dorpe}}]{Shehata2023}%
  \BibitemOpen
  \bibfield  {author} {\bibinfo {author} {\bibfnamefont {M.~M. E.~K.}\
  \bibnamefont {Shehata}}, \bibinfo {author} {\bibfnamefont {G.}~\bibnamefont
  {Simion}}, \bibinfo {author} {\bibfnamefont {R.}~\bibnamefont {Li}}, \bibinfo
  {author} {\bibfnamefont {F.~A.}\ \bibnamefont {Mohiyaddin}}, \bibinfo
  {author} {\bibfnamefont {D.}~\bibnamefont {Wan}}, \bibinfo {author}
  {\bibfnamefont {M.}~\bibnamefont {Mongillo}}, \bibinfo {author}
  {\bibfnamefont {B.}~\bibnamefont {Govoreanu}}, \bibinfo {author}
  {\bibfnamefont {I.}~\bibnamefont {Radu}}, \bibinfo {author} {\bibfnamefont
  {K.}~\bibnamefont {De~Greve}},\ and\ \bibinfo {author} {\bibfnamefont
  {P.}~\bibnamefont {Van~Dorpe}},\ }\bibfield  {title} {\bibinfo {title}
  {Modeling semiconductor spin qubits and their charge noise environment for
  quantum gate fidelity estimation},\ }\href
  {https://doi.org/10.1103/PhysRevB.108.045305} {\bibfield  {journal} {\bibinfo
   {journal} {Phys. Rev. B}\ }\textbf {\bibinfo {volume} {108}},\ \bibinfo
  {pages} {045305} (\bibinfo {year} {2023})}\BibitemShut {NoStop}%
\bibitem [{\citenamefont {Beaudoin}\ and\ \citenamefont
  {Coish}(2015)}]{Beaudoin2015}%
  \BibitemOpen
  \bibfield  {author} {\bibinfo {author} {\bibfnamefont {F.}~\bibnamefont
  {Beaudoin}}\ and\ \bibinfo {author} {\bibfnamefont {W.~A.}\ \bibnamefont
  {Coish}},\ }\bibfield  {title} {\bibinfo {title} {Microscopic models for
  charge-noise-induced dephasing of solid-state qubits},\ }\href
  {https://doi.org/10.1103/PhysRevB.91.165432} {\bibfield  {journal} {\bibinfo
  {journal} {Phys. Rev. B}\ }\textbf {\bibinfo {volume} {91}},\ \bibinfo
  {pages} {165432} (\bibinfo {year} {2015})}\BibitemShut {NoStop}%
\bibitem [{\citenamefont {Kepa}\ \emph
  {et~al.}(2023{\natexlab{a}})\citenamefont {Kepa}, \citenamefont {Focke},
  \citenamefont {Cywinski},\ and\ \citenamefont {Krzywda}}]{Kepa2023}%
  \BibitemOpen
  \bibfield  {author} {\bibinfo {author} {\bibfnamefont {M.}~\bibnamefont
  {Kepa}}, \bibinfo {author} {\bibfnamefont {N.}~\bibnamefont {Focke}},
  \bibinfo {author} {\bibfnamefont {L.}~\bibnamefont {Cywinski}},\ and\
  \bibinfo {author} {\bibfnamefont {J.}~\bibnamefont {Krzywda}},\ }\bibfield
  {title} {\bibinfo {title} {Simulation of 1/f charge noise affecting a quantum
  dot in a {Si/SiGe} structure},\ }\href {https://doi.org/10.1063/5.0151029}
  {\bibfield  {journal} {\bibinfo  {journal} {Applied Physics Letters}\
  }\textbf {\bibinfo {volume} {123}},\ \bibinfo {pages} {034005} (\bibinfo
  {year} {2023}{\natexlab{a}})}\BibitemShut {NoStop}%
\bibitem [{\citenamefont {Kepa}\ \emph
  {et~al.}(2023{\natexlab{b}})\citenamefont {Kepa}, \citenamefont {Cywinski},\
  and\ \citenamefont {Krzywda}}]{KepaCorrelations2023}%
  \BibitemOpen
  \bibfield  {author} {\bibinfo {author} {\bibfnamefont {M.}~\bibnamefont
  {Kepa}}, \bibinfo {author} {\bibfnamefont {L.}~\bibnamefont {Cywinski}},\
  and\ \bibinfo {author} {\bibfnamefont {J.~A.}\ \bibnamefont {Krzywda}},\
  }\bibfield  {title} {\bibinfo {title} {Correlations of spin splitting and
  orbital fluctuations due to 1/f charge noise in the {Si/SiGe} quantum dot},\
  }\href {https://doi.org/10.1063/5.0156358} {\bibfield  {journal} {\bibinfo
  {journal} {Applied Physics Letters}\ }\textbf {\bibinfo {volume} {123}},\
  \bibinfo {pages} {034003} (\bibinfo {year} {2023}{\natexlab{b}})}\BibitemShut
  {NoStop}%
\bibitem [{\citenamefont {King}\ \emph {et~al.}(2020)\citenamefont {King},
  \citenamefont {Schoenfield}, \citenamefont {Calder{\'o}n}, \citenamefont
  {Koiller}, \citenamefont {Saraiva}, \citenamefont {Hu}, \citenamefont
  {Jiang}, \citenamefont {Friesen},\ and\ \citenamefont
  {Coppersmith}}]{King2020}%
  \BibitemOpen
  \bibfield  {author} {\bibinfo {author} {\bibfnamefont {C.}~\bibnamefont
  {King}}, \bibinfo {author} {\bibfnamefont {J.~S.}\ \bibnamefont
  {Schoenfield}}, \bibinfo {author} {\bibfnamefont {M.~J.}\ \bibnamefont
  {Calder{\'o}n}}, \bibinfo {author} {\bibfnamefont {B.}~\bibnamefont
  {Koiller}}, \bibinfo {author} {\bibfnamefont {A.}~\bibnamefont {Saraiva}},
  \bibinfo {author} {\bibfnamefont {X.}~\bibnamefont {Hu}}, \bibinfo {author}
  {\bibfnamefont {H.~W.}\ \bibnamefont {Jiang}}, \bibinfo {author}
  {\bibfnamefont {M.}~\bibnamefont {Friesen}},\ and\ \bibinfo {author}
  {\bibfnamefont {S.~N.}\ \bibnamefont {Coppersmith}},\ }\bibfield  {title}
  {\bibinfo {title} {Lifting of spin blockade by charged impurities in {Si-MOS}
  double quantum dot devices},\ }\href@noop {} {\bibfield  {journal} {\bibinfo
  {journal} {Phys. Rev. B}\ }\textbf {\bibinfo {volume} {101}},\ \bibinfo
  {pages} {155411} (\bibinfo {year} {2020})}\BibitemShut {NoStop}%
\bibitem [{\citenamefont {Yu}(1985)}]{Yu1985}%
  \BibitemOpen
  \bibfield  {author} {\bibinfo {author} {\bibfnamefont {C.~C.}\ \bibnamefont
  {Yu}},\ }\bibfield  {title} {\bibinfo {title} {Possible mechanism for thermal
  conductivity in {(KBr)$_{1\text{-}x}$(KCN)$_x$}},\ }\href@noop {} {\bibfield
  {journal} {\bibinfo  {journal} {Phys. Rev. B}\ }\textbf {\bibinfo {volume}
  {32}},\ \bibinfo {pages} {4220} (\bibinfo {year} {1985})}\BibitemShut
  {NoStop}%
\bibitem [{\citenamefont {Fowler}\ \emph {et~al.}(2012)\citenamefont {Fowler},
  \citenamefont {Mariantoni}, \citenamefont {Martinis},\ and\ \citenamefont
  {Cleland}}]{Fowler2012}%
  \BibitemOpen
  \bibfield  {author} {\bibinfo {author} {\bibfnamefont {A.~G.}\ \bibnamefont
  {Fowler}}, \bibinfo {author} {\bibfnamefont {M.}~\bibnamefont {Mariantoni}},
  \bibinfo {author} {\bibfnamefont {J.~M.}\ \bibnamefont {Martinis}},\ and\
  \bibinfo {author} {\bibfnamefont {A.~N.}\ \bibnamefont {Cleland}},\
  }\bibfield  {title} {\bibinfo {title} {Surface codes: Towards practical
  large-scale quantum computation},\ }\href
  {https://doi.org/10.1103/PhysRevA.86.032324} {\bibfield  {journal} {\bibinfo
  {journal} {Phys. Rev. A}\ }\textbf {\bibinfo {volume} {86}},\ \bibinfo
  {pages} {032324} (\bibinfo {year} {2012})}\BibitemShut {NoStop}%
\bibitem [{\citenamefont {Hunklinger}\ and\ \citenamefont
  {Raychaudhuri}(1986)}]{Hunklinger1986}%
  \BibitemOpen
  \bibfield  {author} {\bibinfo {author} {\bibfnamefont {S.}~\bibnamefont
  {Hunklinger}}\ and\ \bibinfo {author} {\bibfnamefont {A.~K.}\ \bibnamefont
  {Raychaudhuri}},\ }\bibfield  {title} {\bibinfo {title} {Thermal and elastic
  anomalies in glasses at low temperatures},\ }\href@noop {} {\bibfield
  {journal} {\bibinfo  {journal} {Prog. in Low Temp. Phys.}\ }\textbf {\bibinfo
  {volume} {9}},\ \bibinfo {pages} {265} (\bibinfo {year} {1986})}\BibitemShut
  {NoStop}%
\bibitem [{\citenamefont {Phillips}(1987)}]{Phillips1987}%
  \BibitemOpen
  \bibfield  {author} {\bibinfo {author} {\bibfnamefont {W.~A.}\ \bibnamefont
  {Phillips}},\ }\bibfield  {title} {\bibinfo {title} {Two-level states in
  glasses},\ }\href@noop {} {\bibfield  {journal} {\bibinfo  {journal} {Rep.
  Prog. Phys.}\ }\textbf {\bibinfo {volume} {50}},\ \bibinfo {pages} {1657}
  (\bibinfo {year} {1987})}\BibitemShut {NoStop}%
\bibitem [{\citenamefont {Anderson}\ \emph {et~al.}(1972)\citenamefont
  {Anderson}, \citenamefont {Halperin},\ and\ \citenamefont
  {Varma}}]{Anderson1972}%
  \BibitemOpen
  \bibfield  {author} {\bibinfo {author} {\bibfnamefont {P.~W.}\ \bibnamefont
  {Anderson}}, \bibinfo {author} {\bibfnamefont {B.~I.}\ \bibnamefont
  {Halperin}},\ and\ \bibinfo {author} {\bibfnamefont {C.~M.}\ \bibnamefont
  {Varma}},\ }\bibfield  {title} {\bibinfo {title} {Anomalous low-temperature
  thermal properties of glasses and spin glasses},\ }\href
  {https://doi.org/10.1080/14786437208229210} {\bibfield  {journal} {\bibinfo
  {journal} {The Philosophical Magazine: A Journal of Theoretical Experimental
  and Applied Physics}\ }\textbf {\bibinfo {volume} {25}},\ \bibinfo {pages}
  {1} (\bibinfo {year} {1972})},\ \Eprint
  {https://arxiv.org/abs/https://doi.org/10.1080/14786437208229210}
  {https://doi.org/10.1080/14786437208229210} \BibitemShut {NoStop}%
\bibitem [{\citenamefont {Phillips}(1972)}]{Phillips1972}%
  \BibitemOpen
  \bibfield  {author} {\bibinfo {author} {\bibfnamefont {W.~A.}\ \bibnamefont
  {Phillips}},\ }\bibfield  {title} {\bibinfo {title} {Tunneling states in
  amorphous solids},\ }\href@noop {} {\bibfield  {journal} {\bibinfo  {journal}
  {Journal of Low Temperature Physics}\ }\textbf {\bibinfo {volume} {7}},\
  \bibinfo {pages} {351} (\bibinfo {year} {1972})}\BibitemShut {NoStop}%
\bibitem [{\citenamefont {Carruzzo}\ and\ \citenamefont
  {Yu}(2020)}]{Carruzzo2020}%
  \BibitemOpen
  \bibfield  {author} {\bibinfo {author} {\bibfnamefont {H.~M.}\ \bibnamefont
  {Carruzzo}}\ and\ \bibinfo {author} {\bibfnamefont {C.~C.}\ \bibnamefont
  {Yu}},\ }\bibfield  {title} {\bibinfo {title} {Why phonon scattering in
  glasses is universally small at low temperatures},\ }\href
  {https://doi.org/10.1103/PhysRevLett.124.075902} {\bibfield  {journal}
  {\bibinfo  {journal} {Phys. Rev. Lett.}\ }\textbf {\bibinfo {volume} {124}},\
  \bibinfo {pages} {075902} (\bibinfo {year} {2020})}\BibitemShut {NoStop}%
\bibitem [{\citenamefont {Yu}\ and\ \citenamefont {Leggett}(1988)}]{Yu1988}%
  \BibitemOpen
  \bibfield  {author} {\bibinfo {author} {\bibfnamefont {C.~C.}\ \bibnamefont
  {Yu}}\ and\ \bibinfo {author} {\bibfnamefont {A.~J.}\ \bibnamefont
  {Leggett}},\ }\bibfield  {title} {\bibinfo {title} {Low temperature
  properties of amorphous materials: Through a glass darkly},\ }\href@noop {}
  {\bibfield  {journal} {\bibinfo  {journal} {Comments Cond. Mat. Phys.}\
  }\textbf {\bibinfo {volume} {14}},\ \bibinfo {pages} {231} (\bibinfo {year}
  {1988})}\BibitemShut {NoStop}%
\bibitem [{\citenamefont {Salvino}\ \emph {et~al.}(1994)\citenamefont
  {Salvino}, \citenamefont {Rogge}, \citenamefont {Tigner},\ and\ \citenamefont
  {Osheroff}}]{Salvino1994}%
  \BibitemOpen
  \bibfield  {author} {\bibinfo {author} {\bibfnamefont {D.~J.}\ \bibnamefont
  {Salvino}}, \bibinfo {author} {\bibfnamefont {S.}~\bibnamefont {Rogge}},
  \bibinfo {author} {\bibfnamefont {B.}~\bibnamefont {Tigner}},\ and\ \bibinfo
  {author} {\bibfnamefont {D.~D.}\ \bibnamefont {Osheroff}},\ }\bibfield
  {title} {\bibinfo {title} {Low temperature {AC} dielectric response of
  glasses to high {DC} electric fields},\ }\href@noop {} {\bibfield  {journal}
  {\bibinfo  {journal} {Phys. Rev. Lett.}\ }\textbf {\bibinfo {volume} {73}},\
  \bibinfo {pages} {268} (\bibinfo {year} {1994})}\BibitemShut {NoStop}%
\bibitem [{\citenamefont {Carruzzo}\ \emph {et~al.}(1994)\citenamefont
  {Carruzzo}, \citenamefont {Grannan},\ and\ \citenamefont
  {Yu}}]{Carruzzo1994}%
  \BibitemOpen
  \bibfield  {author} {\bibinfo {author} {\bibfnamefont {H.~M.}\ \bibnamefont
  {Carruzzo}}, \bibinfo {author} {\bibfnamefont {E.~R.}\ \bibnamefont
  {Grannan}},\ and\ \bibinfo {author} {\bibfnamefont {C.~C.}\ \bibnamefont
  {Yu}},\ }\bibfield  {title} {\bibinfo {title} {Nonequilibrium dielectric
  behavior in glasses at low temperatures: evidence for interacting defects},\
  }\href@noop {} {\bibfield  {journal} {\bibinfo  {journal} {Phys. Rev. B}\
  }\textbf {\bibinfo {volume} {50}},\ \bibinfo {pages} {6685} (\bibinfo {year}
  {1994})}\BibitemShut {NoStop}%
\bibitem [{\citenamefont {Joffrin}\ and\ \citenamefont
  {Levelut}(1975)}]{Joffrin1975}%
  \BibitemOpen
  \bibfield  {author} {\bibinfo {author} {\bibfnamefont {J.}~\bibnamefont
  {Joffrin}}\ and\ \bibinfo {author} {\bibfnamefont {A.}~\bibnamefont
  {Levelut}},\ }\bibfield  {title} {\bibinfo {title} {Virtual phonon exchange
  in glasses},\ }\href {http://dx.doi.org/10.1051/jphys:01975003609081100}
  {\bibfield  {journal} {\bibinfo  {journal} {J. Phys. France}\ }\textbf
  {\bibinfo {volume} {36}} (\bibinfo {year} {1975})}\BibitemShut {NoStop}%
\bibitem [{\citenamefont {Berret}\ and\ \citenamefont
  {Mei{\ss}ner}(1988)}]{Berret1988}%
  \BibitemOpen
  \bibfield  {author} {\bibinfo {author} {\bibfnamefont {J.~F.}\ \bibnamefont
  {Berret}}\ and\ \bibinfo {author} {\bibfnamefont {M.}~\bibnamefont
  {Mei{\ss}ner}},\ }\bibfield  {title} {\bibinfo {title} {How universal are the
  low temperature acoustic properties of glasses?},\ }\href
  {https://doi.org/10.1007/BF01320540} {\bibfield  {journal} {\bibinfo
  {journal} {Z. Phys. B}\ }\textbf {\bibinfo {volume} {70}},\ \bibinfo {pages}
  {65} (\bibinfo {year} {1988})}\BibitemShut {NoStop}%
\bibitem [{\citenamefont {Yu}\ and\ \citenamefont {Freeman}(1987)}]{Yu1987}%
  \BibitemOpen
  \bibfield  {author} {\bibinfo {author} {\bibfnamefont {C.~C.}\ \bibnamefont
  {Yu}}\ and\ \bibinfo {author} {\bibfnamefont {J.~J.}\ \bibnamefont
  {Freeman}},\ }\bibfield  {title} {\bibinfo {title} {Thermal conductivity and
  specific heat of glasses},\ }\href {https://doi.org/10.1103/PhysRevB.36.7620}
  {\bibfield  {journal} {\bibinfo  {journal} {Phys. Rev. B}\ }\textbf {\bibinfo
  {volume} {36}},\ \bibinfo {pages} {7620} (\bibinfo {year}
  {1987})}\BibitemShut {NoStop}%
\bibitem [{\citenamefont {Golding}\ \emph {et~al.}(1979)\citenamefont
  {Golding}, \citenamefont {Schickfus}, \citenamefont {Hunklinger},\ and\
  \citenamefont {Dransfeld}}]{Golding1979}%
  \BibitemOpen
  \bibfield  {author} {\bibinfo {author} {\bibfnamefont {B.}~\bibnamefont
  {Golding}}, \bibinfo {author} {\bibfnamefont {M.~v.}\ \bibnamefont
  {Schickfus}}, \bibinfo {author} {\bibfnamefont {S.}~\bibnamefont
  {Hunklinger}},\ and\ \bibinfo {author} {\bibfnamefont {K.}~\bibnamefont
  {Dransfeld}},\ }\bibfield  {title} {\bibinfo {title} {Intrinsic electric
  dipole moment of tunneling systems in silica glasses},\ }\href
  {https://doi.org/10.1103/PhysRevLett.43.1817} {\bibfield  {journal} {\bibinfo
   {journal} {Phys. Rev. Lett.}\ }\textbf {\bibinfo {volume} {43}},\ \bibinfo
  {pages} {1817} (\bibinfo {year} {1979})}\BibitemShut {NoStop}%
\bibitem [{\citenamefont {Kogan}(1996)}]{Kogan1996}%
  \BibitemOpen
  \bibfield  {author} {\bibinfo {author} {\bibfnamefont {S.}~\bibnamefont
  {Kogan}},\ }\href@noop {} {\emph {\bibinfo {title} {Electronic Noise and
  Fluctuations in Solids}}}\ (\bibinfo  {publisher} {Cambridge University
  Press},\ \bibinfo {address} {Cambridge},\ \bibinfo {year} {1996})\BibitemShut
  {NoStop}%
\bibitem [{\citenamefont {Faoro}\ and\ \citenamefont
  {Ioffe}(2006)}]{Faoro2006}%
  \BibitemOpen
  \bibfield  {author} {\bibinfo {author} {\bibfnamefont {L.}~\bibnamefont
  {Faoro}}\ and\ \bibinfo {author} {\bibfnamefont {L.~B.}\ \bibnamefont
  {Ioffe}},\ }\bibfield  {title} {\bibinfo {title} {Quantum two level systems
  and kondo-like traps as possible sources of decoherence in superconducting
  qubits},\ }\href {https://doi.org/10.1103/PhysRevLett.96.047001} {\bibfield
  {journal} {\bibinfo  {journal} {Phys. Rev. Lett.}\ }\textbf {\bibinfo
  {volume} {96}},\ \bibinfo {pages} {047001} (\bibinfo {year}
  {2006})}\BibitemShut {NoStop}%
\bibitem [{\citenamefont {Constantin}\ \emph {et~al.}(2009)\citenamefont
  {Constantin}, \citenamefont {Yu},\ and\ \citenamefont
  {Martinis}}]{Constantin2009}%
  \BibitemOpen
  \bibfield  {author} {\bibinfo {author} {\bibfnamefont {M.}~\bibnamefont
  {Constantin}}, \bibinfo {author} {\bibfnamefont {C.~C.}\ \bibnamefont {Yu}},\
  and\ \bibinfo {author} {\bibfnamefont {J.~M.}\ \bibnamefont {Martinis}},\
  }\bibfield  {title} {\bibinfo {title} {Saturation of two-level systems and
  charge noise in josephson junction qubits},\ }\href
  {https://doi.org/10.1103/PhysRevB.79.094520} {\bibfield  {journal} {\bibinfo
  {journal} {Phys. Rev. B}\ }\textbf {\bibinfo {volume} {79}},\ \bibinfo
  {pages} {094520} (\bibinfo {year} {2009})}\BibitemShut {NoStop}%
\bibitem [{\citenamefont {Metropolis}\ \emph {et~al.}(1953)\citenamefont
  {Metropolis}, \citenamefont {Rosenbluth}, \citenamefont {Rosenbluth},
  \citenamefont {Teller},\ and\ \citenamefont {Teller}}]{Metropolis1953}%
  \BibitemOpen
  \bibfield  {author} {\bibinfo {author} {\bibfnamefont {N.}~\bibnamefont
  {Metropolis}}, \bibinfo {author} {\bibfnamefont {A.~W.}\ \bibnamefont
  {Rosenbluth}}, \bibinfo {author} {\bibfnamefont {M.~N.}\ \bibnamefont
  {Rosenbluth}}, \bibinfo {author} {\bibfnamefont {A.~H.}\ \bibnamefont
  {Teller}},\ and\ \bibinfo {author} {\bibfnamefont {E.}~\bibnamefont
  {Teller}},\ }\bibfield  {title} {\bibinfo {title} {Equation of state
  calculations by fast computing machines},\ }\href
  {https://doi.org/10.1063/1.1699114} {\bibfield  {journal} {\bibinfo
  {journal} {J. Chem. Phys.}\ }\textbf {\bibinfo {volume} {21}},\ \bibinfo
  {pages} {1087} (\bibinfo {year} {1953})}\BibitemShut {NoStop}%
\bibitem [{\citenamefont {Bhatt}\ and\ \citenamefont
  {Young}(1988)}]{Bhatt1988}%
  \BibitemOpen
  \bibfield  {author} {\bibinfo {author} {\bibfnamefont {R.~N.}\ \bibnamefont
  {Bhatt}}\ and\ \bibinfo {author} {\bibfnamefont {A.~P.}\ \bibnamefont
  {Young}},\ }\bibfield  {title} {\bibinfo {title} {Numerical studies of ising
  spin glasses in two, three, and four dimensions},\ }\href
  {https://doi.org/10.1103/PhysRevB.37.5606} {\bibfield  {journal} {\bibinfo
  {journal} {Phys. Rev. B}\ }\textbf {\bibinfo {volume} {37}},\ \bibinfo
  {pages} {5606} (\bibinfo {year} {1988})}\BibitemShut {NoStop}%
\bibitem [{\citenamefont {Press}\ \emph {et~al.}(1992)\citenamefont {Press},
  \citenamefont {Teukolsky}, \citenamefont {Vetterling},\ and\ \citenamefont
  {Flannery}}]{Press1992}%
  \BibitemOpen
  \bibfield  {author} {\bibinfo {author} {\bibfnamefont {W.~H.}\ \bibnamefont
  {Press}}, \bibinfo {author} {\bibfnamefont {S.~A.}\ \bibnamefont
  {Teukolsky}}, \bibinfo {author} {\bibfnamefont {W.~T.}\ \bibnamefont
  {Vetterling}},\ and\ \bibinfo {author} {\bibfnamefont {B.~P.}\ \bibnamefont
  {Flannery}},\ }\href@noop {} {\emph {\bibinfo {title} {Numerical Recipes in
  C: The Art of Scientific Computing}}}\ (\bibinfo  {publisher} {Cambridge
  University Press},\ \bibinfo {address} {New York},\ \bibinfo {year}
  {1992})\BibitemShut {NoStop}%
\bibitem [{\citenamefont {Frigo}\ and\ \citenamefont
  {Johnson}(2005)}]{Frigo2005}%
  \BibitemOpen
  \bibfield  {author} {\bibinfo {author} {\bibfnamefont {M.}~\bibnamefont
  {Frigo}}\ and\ \bibinfo {author} {\bibfnamefont {S.~G.}\ \bibnamefont
  {Johnson}},\ }\bibfield  {title} {\bibinfo {title} {The design and
  implementation of {FFTW3}},\ }\href@noop {} {\bibfield  {journal} {\bibinfo
  {journal} {Proceedings of the IEEE}\ }\textbf {\bibinfo {volume} {93}},\
  \bibinfo {pages} {216} (\bibinfo {year} {2005})},\ \bibinfo {note} {special
  issue on ``Program Generation, Optimization, and Platform
  Adaptation''}\BibitemShut {NoStop}%
\bibitem [{\citenamefont {Jock}\ \emph {et~al.}(2022)\citenamefont {Jock},
  \citenamefont {Jacobson}, \citenamefont {Rudolph}, \citenamefont {Ward},
  \citenamefont {Carroll},\ and\ \citenamefont {Luhman}}]{Jock2022}%
  \BibitemOpen
  \bibfield  {author} {\bibinfo {author} {\bibfnamefont {R.~M.}\ \bibnamefont
  {Jock}}, \bibinfo {author} {\bibfnamefont {N.~T.}\ \bibnamefont {Jacobson}},
  \bibinfo {author} {\bibfnamefont {M.}~\bibnamefont {Rudolph}}, \bibinfo
  {author} {\bibfnamefont {D.~R.}\ \bibnamefont {Ward}}, \bibinfo {author}
  {\bibfnamefont {M.~S.}\ \bibnamefont {Carroll}},\ and\ \bibinfo {author}
  {\bibfnamefont {D.~R.}\ \bibnamefont {Luhman}},\ }\bibfield  {title}
  {\bibinfo {title} {A silicon singlet-triplet qubit driven by spin-valley
  coupling},\ }\href {https://doi.org/10.1038/s41467-022-28302-y} {\bibfield
  {journal} {\bibinfo  {journal} {Nature Comm.}\ }\textbf {\bibinfo {volume}
  {13}},\ \bibinfo {pages} {641} (\bibinfo {year} {2022})}\BibitemShut
  {NoStop}%
\bibitem [{\citenamefont {Ogielski}(1985)}]{Ogielski1985}%
  \BibitemOpen
  \bibfield  {author} {\bibinfo {author} {\bibfnamefont {A.~T.}\ \bibnamefont
  {Ogielski}},\ }\bibfield  {title} {\bibinfo {title} {Dynamics of
  three-dimensional ising spin glasses in thermal equilibrium},\ }\href
  {https://doi.org/10.1103/PhysRevB.32.7384} {\bibfield  {journal} {\bibinfo
  {journal} {Phys. Rev. B}\ }\textbf {\bibinfo {volume} {32}},\ \bibinfo
  {pages} {7384} (\bibinfo {year} {1985})}\BibitemShut {NoStop}%
\bibitem [{\citenamefont {Snider}\ and\ \citenamefont {Yu}(2005)}]{Snider2005}%
  \BibitemOpen
  \bibfield  {author} {\bibinfo {author} {\bibfnamefont {J.}~\bibnamefont
  {Snider}}\ and\ \bibinfo {author} {\bibfnamefont {C.~C.}\ \bibnamefont
  {Yu}},\ }\bibfield  {title} {\bibinfo {title} {Absence of dipole glass
  transition for randomly dilute classical ising dipoles},\ }\href
  {https://doi.org/10.1103/PhysRevB.72.214203} {\bibfield  {journal} {\bibinfo
  {journal} {Phys. Rev. B}\ }\textbf {\bibinfo {volume} {72}},\ \bibinfo
  {pages} {214203} (\bibinfo {year} {2005})}\BibitemShut {NoStop}%
\bibitem [{\citenamefont {Quilliam}\ \emph {et~al.}(2007)\citenamefont
  {Quilliam}, \citenamefont {Mugford}, \citenamefont {Gomez}, \citenamefont
  {Kycia},\ and\ \citenamefont {Kycia}}]{Quilliam2007}%
  \BibitemOpen
  \bibfield  {author} {\bibinfo {author} {\bibfnamefont {J.~A.}\ \bibnamefont
  {Quilliam}}, \bibinfo {author} {\bibfnamefont {C.~G.~A.}\ \bibnamefont
  {Mugford}}, \bibinfo {author} {\bibfnamefont {A.}~\bibnamefont {Gomez}},
  \bibinfo {author} {\bibfnamefont {S.~W.}\ \bibnamefont {Kycia}},\ and\
  \bibinfo {author} {\bibfnamefont {J.~B.}\ \bibnamefont {Kycia}},\ }\bibfield
  {title} {\bibinfo {title} {Specific heat of the dilute ising magnet
  {${\mathrm{LiHo}}_{x}{\mathrm{Y}}_{1\ensuremath{-}x}{\mathrm{F}}_{4}$}},\
  }\href {https://doi.org/10.1103/PhysRevLett.98.037203} {\bibfield  {journal}
  {\bibinfo  {journal} {Phys. Rev. Lett.}\ }\textbf {\bibinfo {volume} {98}},\
  \bibinfo {pages} {037203} (\bibinfo {year} {2007})}\BibitemShut {NoStop}%
\bibitem [{\citenamefont {J\"onsson}\ \emph {et~al.}(2007)\citenamefont
  {J\"onsson}, \citenamefont {Mathieu}, \citenamefont {Wernsdorfer},
  \citenamefont {Tkachuk},\ and\ \citenamefont {Barbara}}]{Jonsson2007}%
  \BibitemOpen
  \bibfield  {author} {\bibinfo {author} {\bibfnamefont {P.~E.}\ \bibnamefont
  {J\"onsson}}, \bibinfo {author} {\bibfnamefont {R.}~\bibnamefont {Mathieu}},
  \bibinfo {author} {\bibfnamefont {W.}~\bibnamefont {Wernsdorfer}}, \bibinfo
  {author} {\bibfnamefont {A.~M.}\ \bibnamefont {Tkachuk}},\ and\ \bibinfo
  {author} {\bibfnamefont {B.}~\bibnamefont {Barbara}},\ }\bibfield  {title}
  {\bibinfo {title} {Absence of conventional spin-glass transition in the ising
  dipolar system
  {${\mathrm{LiHo}}_{x}{\mathrm{Y}}_{1\ensuremath{-}x}{\mathrm{F}}_{4}$}},\
  }\href {https://doi.org/10.1103/PhysRevLett.98.256403} {\bibfield  {journal}
  {\bibinfo  {journal} {Phys. Rev. Lett.}\ }\textbf {\bibinfo {volume} {98}},\
  \bibinfo {pages} {256403} (\bibinfo {year} {2007})}\BibitemShut {NoStop}%
\bibitem [{\citenamefont {Chen}\ and\ \citenamefont {Yu}(2007)}]{Chen2007}%
  \BibitemOpen
  \bibfield  {author} {\bibinfo {author} {\bibfnamefont {Z.}~\bibnamefont
  {Chen}}\ and\ \bibinfo {author} {\bibfnamefont {C.~C.}\ \bibnamefont {Yu}},\
  }\bibfield  {title} {\bibinfo {title} {Measurement-noise maximum as a
  signature of a phase transition},\ }\href
  {https://doi.org/10.1103/PhysRevLett.98.057204} {\bibfield  {journal}
  {\bibinfo  {journal} {Phys. Rev. Lett.}\ }\textbf {\bibinfo {volume} {98}},\
  \bibinfo {pages} {057204} (\bibinfo {year} {2007})}\BibitemShut {NoStop}%
\bibitem [{\citenamefont {Zaitlin}\ and\ \citenamefont
  {Anderson}(1975)}]{Zaitlin1975}%
  \BibitemOpen
  \bibfield  {author} {\bibinfo {author} {\bibfnamefont {M.~P.}\ \bibnamefont
  {Zaitlin}}\ and\ \bibinfo {author} {\bibfnamefont {A.~C.}\ \bibnamefont
  {Anderson}},\ }\bibfield  {title} {\bibinfo {title} {Phonon thermal transport
  in noncrystalline materials},\ }\href
  {https://doi.org/10.1103/PhysRevB.12.4475} {\bibfield  {journal} {\bibinfo
  {journal} {Phys. Rev. B}\ }\textbf {\bibinfo {volume} {12}},\ \bibinfo
  {pages} {4475} (\bibinfo {year} {1975})}\BibitemShut {NoStop}%
\bibitem [{\citenamefont {Black}(1978)}]{Black1978}%
  \BibitemOpen
  \bibfield  {author} {\bibinfo {author} {\bibfnamefont {J.~L.}\ \bibnamefont
  {Black}},\ }\bibfield  {title} {\bibinfo {title} {Relationship between the
  time-dependent specific heat and the ultrasonic properties of glasses at low
  temperatures},\ }\href {https://doi.org/10.1103/PhysRevB.17.2740} {\bibfield
  {journal} {\bibinfo  {journal} {Phys. Rev. B}\ }\textbf {\bibinfo {volume}
  {17}},\ \bibinfo {pages} {2740} (\bibinfo {year} {1978})}\BibitemShut
  {NoStop}%
\bibitem [{\citenamefont {Zimmermann}\ and\ \citenamefont
  {Weber}(1981)}]{Zimmermann1981}%
  \BibitemOpen
  \bibfield  {author} {\bibinfo {author} {\bibfnamefont {J.}~\bibnamefont
  {Zimmermann}}\ and\ \bibinfo {author} {\bibfnamefont {G.}~\bibnamefont
  {Weber}},\ }\bibfield  {title} {\bibinfo {title} {Thermal relaxation of
  low-energy excitations in vitreous silica},\ }\href
  {https://doi.org/10.1103/PhysRevLett.46.661} {\bibfield  {journal} {\bibinfo
  {journal} {Phys. Rev. Lett.}\ }\textbf {\bibinfo {volume} {46}},\ \bibinfo
  {pages} {661} (\bibinfo {year} {1981})}\BibitemShut {NoStop}%
\bibitem [{\citenamefont {Yu}\ and\ \citenamefont {Carruzzo}(2004)}]{Yu2004}%
  \BibitemOpen
  \bibfield  {author} {\bibinfo {author} {\bibfnamefont {C.~C.}\ \bibnamefont
  {Yu}}\ and\ \bibinfo {author} {\bibfnamefont {H.~M.}\ \bibnamefont
  {Carruzzo}},\ }\bibfield  {title} {\bibinfo {title} {Frequency dependence and
  equilibration of the specific heat of glass-forming liquids},\ }\href
  {https://doi.org/10.1103/PhysRevE.69.051201} {\bibfield  {journal} {\bibinfo
  {journal} {Phys. Rev. E}\ }\textbf {\bibinfo {volume} {69}},\ \bibinfo
  {pages} {051201} (\bibinfo {year} {2004})}\BibitemShut {NoStop}%
\end{thebibliography}

%
\end{document}